\documentclass[journal]{IEEEtran}
\usepackage{amsmath,amssymb,amsfonts,graphicx,epsfig,cite,fancyhdr,amsthm,graphicx}
\usepackage{dsfont, color}
\usepackage{subfigure}
\usepackage[center]{caption}
\usepackage{chngpage}
\usepackage{multicol}
\usepackage{array}
\usepackage{algorithmic}
\usepackage{multirow, rotating}
\usepackage[square, comma, numbers, sort&compress]{natbib}
\usepackage{epstopdf}

\usepackage{tikz}
\usetikzlibrary{mindmap,trees,matrix, arrows,shapes,positioning,shadows}
\usetikzlibrary{decorations.pathmorphing} 
\usetikzlibrary{fit,chains, calc}					
\usetikzlibrary{backgrounds}	
\definecolor{myblue}{RGB}{50,150,191}
\definecolor{OliveGreen}{RGB}{141, 182, 0}

\theoremstyle{definition}

\newtheorem{result}{Result}

\title{\fontsize{23}{28}\selectfont Impact of Propagation Environment on Energy-Efficient Relay Placement: Model and Performance Analysis}
\author{Fanny Parzysz, \IEEEmembership{Student Member IEEE}, Mai Vu, \IEEEmembership{Senior Member IEEE}, Fran\c cois Gagnon, \IEEEmembership{Senior Member IEEE}%
\thanks{This work has been supported in part by Ultra Electronics TCS and the Natural Science and Engineering Council of Canada as part of the “High Performance Emergency and Tactical Wireless Communication Chair” at \'{E}cole de Technologie Sup\'{e}rieure.}
\thanks{F. Parzysz and F. Gagnon are with \'Ecole de Technologie Sup\'erieure, Montreal, Canada. M. Vu is with the  Electrical and Computer Engineering department at Tufts University. (emails: Fanny.Parzysz@lacime.etsmtl.ca; Mai.Vu@tufts.edu; Francois.Gagnon@etsmtl.ca) }
}

\begin{document}
\maketitle

\newcounter{algorithmcounter}
\setcounter{algorithmcounter}{\value{table}}
\addtocounter{algorithmcounter}{1}

\newcounter{tableau}
\setcounter{tableau}{1}

\begin{abstract}
The performance of a relay-based cellular network is greatly affected by the relay location within a cell.
Existing results for optimal relay placement do not reflect how the radio propagation environment and choice of the coding scheme can impact system performance.
In this paper, we analyze the impact on relaying performance of node distances, relay height and line-of-sight conditions for both uplink and downlink transmissions, using several relay coding schemes. 
Our first objective is to propose a geometrical model for energy-efficient relay placement that requires
only a small number of characteristic distances.
Our second objective is to estimate the maximum cell coverage of a relay-aided cell given power constraints, and conversely, the averaged energy consumption given a cell radius.
We show that the practical full decode-forward scheme performs close to the energy-optimized partial decode-forward scheme when the relay is ideally located. However, away from this optimum relay location, performance rapidly degrades and more advanced coding scheme, such as partial decode-forward, is needed to maintain good performance and allow more freedom in the relay placement.
Finally, we define a trade-off between cell coverage and energy efficiency, and show that there exists a relay location for which increasing the cell coverage has a minimal impact on the average energy consumed per unit area.

\end{abstract}

\bstctlcite{IEEEexample:BSTcontrol}

\section{Introduction}
\label{introduction}

Relay-aided transmissions promise significant gains in both coverage extension and performance enhancement.  Relay-based architecture for cellular network has been considered for next generation wireless systems, such as the LTE-A \cite{ghosh2010, loa2010, damnjanovic2011}, 
and is envisioned as part of future heterogeneous networks, along with macro-, pico-, and femtocells. Picocells are operator-deployed and help alleviate coverage dead zones and traffic hot zones, while femtocells are usually consumer-deployed in an unplanned manner to increase in-door coverage. Both are connected to their own wired backhaul connection, in contrary to relays. The use of low power, low complexity relays in lieu of additional wired connected stations avoids expensive backhaul links.
Relay nodes are deployed with a more traditional RF planning, with site deployment and maintenance by the operator, thus allowing optimization of their placement within the cell.
At the same time, as a result of the tremendous increase of subscribers, jointly with the ever-growing demand for application requirements, energy efficiency became a great challenge for future wireless systems. However, the energy impact of the relay location and of the relay coding scheme remains under-explored in the literature.

\subsection{Motivation and Prior work}

Resource allocation and relaying strategies in cooperative systems have mostly targeted improving the cell capacity or spectral efficiency. A rich literature has investigated relay placement with this perspective \cite{lin2007, lichun2008, huang2010, lu2011, sambale2012}. For example, authors of \cite{lin2007} analyze different relaying strategies, amplify-forward, decode-forward and compress-forward, and compare their spectral efficiency.
 
However, the quality of service (QoS) of value-added mobile applications, such as on-demand video, real-time games or video-conferences, is mostly expressed in terms of a minimum rate to be maintained, along with delay and jitter requirements \cite{qos}.
This perspective has spurred new analysis and designs for cellular networks. For example, in \cite{lin2010}, relay placement is optimized to enhance capacity while meeting a traffic demand. With another perspective, references \cite{joshi2011, youssef2012, khakurel2012} address coverage extension, given a user rate. Analysis has also focused on improving cell topology, such as \cite{lin2008} which proposes an interesting novel dual-relay architecture. 

Since user applications mostly run on power-limited devices, reducing the energy consumption, while meeting a rate requirement, has in particular attracted significant attention. Energy minimization for relay-aided cellular networks has ignited new techniques such as relay-based handover strategies \cite{cho2009,yang2009HO}, sleep mode operations \cite{wang2010sleep}, improved resource management \cite{salem2010} and relay-aided load balancing \cite{load_balancing}.
Nevertheless, little can be found on the specific issue of energy-efficient relay placement and on the performance gains obtained by using advanced coding schemes rather than simpler schemes. 
Moreover,
extending cell coverage implies serving far users and consequently, consuming more energy.
This trade-off has neither been investigated.

Another motivation also brought us to analyze energy-efficient relay placement. Although network simulations offer wide possibilities and numerous variable factors for refined investigation, they are often time consuming and may confine the obtained results only to the considered settings. They provide
neither generalization to other settings nor performance limits like an analysis based on capacity bounds.  Thus, a model is needed to give insights of the relay utilization and helps set basis for future investigations and extensive simulations.
Models for heterogeneous networks has attracted significant attention in recent years and various advanced models have been proposed, such as in \cite{Dhillon2012} for femtocell networks. Such models however, are not easy to apply to relay-aided networks, where a cell has to manage both direct and two-hop transmissions and where physical-layer considerations, such as the coding scheme, need to be included.
Current works using models for relaying are mostly based on simplified patterns composed of inner and outer circular regions for direct and relayed transmissions, such as in \cite{Chandwani2010, Schober2011}. In such a model, the radius of the inner region corresponds to the base-station-to-relay distance, which is neither coverage- nor energy-optimized.

Finally, though relay placement has been addressed in the aforementioned works, none of them allows the characterization of the impact of relay placement based on accurate path-loss models. 
First, geometrical models for relay positioning are incomplete. In \cite{yang2010,lin2007},
the considered model is linear, and the relay is located on a straight line between the source and the destination. This model cannot provide accurate solutions for actual relay placement in a network where the user is mobile, while the base station and relay are fixed. In \cite{lichun2008, huang2010, lu2011, yang2011, lin2010, joshi2011, youssef2012, khakurel2012}, two-dimensional models are considered, but the relay height is not taken into account, despite its impact on both path-loss and line-of-sight (LOS) conditions. Three-dimensional models are not fully investigated and relay height has been considered only in \cite{sambale2012}.
Second, a majority of results on relay placement, such as \cite{lin2010, joshi2011, youssef2012, khakurel2012, yang2011,yang2010,lin2007, lu2011}, are for AWGN channels with path-loss exponent, which may be oversimplified as emphasized in \cite{wang2010}, or for fading channels with the same path-loss model for all links in the network.
More accurate path-loss channel models, as in \cite{lichun2008, huang2010, sambale2012}, should be considered.

\subsection{Main contributions}

We introduce in this paper the notion of Relay Efficiency Area and propose a geometrical model for energy-efficient relay placement, for both uplink and downlink transmissions, which allows meaningful performance analyses without the need for excessive simulations.
The proposed model consists of only a few characteristic distances, but accurately approaches simulations on the probability of relaying within 3\% and the energy consumption within 1\%. Then, we build an analysis framework for the performance of relaying based on time-division full decode-forward and provide comparison with the comprehensive partial decode-forward scheme, which is studied with the perspective of rate maximization in \cite{Host-Madsen} and energy minimization in \cite{Journal1}.

In the first part, we consider as a baseline the setting with a fixed number of relays at regular locations  in each cell and no inter-cell interference, leaving these factors to future extensions. This baseline model provides a tractable benchmark for performance comparison and analysis, and shows that the classical inner and outer regions as considered in \cite{Chandwani2010} do not reflect the performance gains obtained by relaying.

In the second part, using the proposed geometrical model, we highlight relay configurations which are beneficial for coverage extension and for energy efficiency. Given that a network designer is constrained by the urban topography and by administrative by-laws for station placement, flexibility in the relay location is also a relevant performance criteria. We thus consider performance both at and nearby the optimal relay location. 
Furthermore, we show that the optimal relay placements for coverage and energy efficiency are different. We thus analyze the trade-off between energy gain and coverage extension and highlight the deployment cost required to reduce the energy consumption. We finally propose a new definition for optimal relay placement regarding the average energy consumed per unit area.

\subsection{Paper overview}

We define our 6-sector urban cell framework and channel path-loss model, in Section \ref{sec:system_model}, and the considered coding schemes in Section \ref{sec:schemes}. Section \ref{sec:model} introduces the characteristic distances that constitute the proposed geometrical model. These distances are derived for both full decode-forward and partial decode-forward in Section \ref{sec:analysis_model}. We then deduce the system performance in Section \ref{sec:P_RTx_energy}.
Simulation results are described in Section \ref{sec:simulation}. Finally, Section \ref{sec:conclusion} concludes this paper.

\textit{Notation:} 
$B$ stands for the base station, $R$ for the relay and $U$ for the user. Calligraphic $\mathcal{R}$ refers to the user rate. Upper-case letters denote constant distances, such as $\mathsf{R}$, $\mathsf{D}$ or $\mathsf{H}$, while lower-case letters stand for variable distances or angles, such as $r$ or $\theta$. Subscripts $_{d}$, $_{s}$ and $_{r}$ respectively refer to the link from user to base station (direct link), user to relay station and relay to base station. $\mathbb{P}$ is used for probabilities and $\mathbb{E}$ for expectation. Finally, DTx refers to direct transmission, while RTx stands for relay-aided transmission, in the considered coding schemes.

\section{System model for a 6-sector urban cell}
\label{sec:system_model}

In this section, we present the cell and path-loss models. In particular, we analyze two different path-loss models, depending on the relay being below or above the rooftop.

\begin{figure*}
\centering \includegraphics[width=0.7\textwidth]{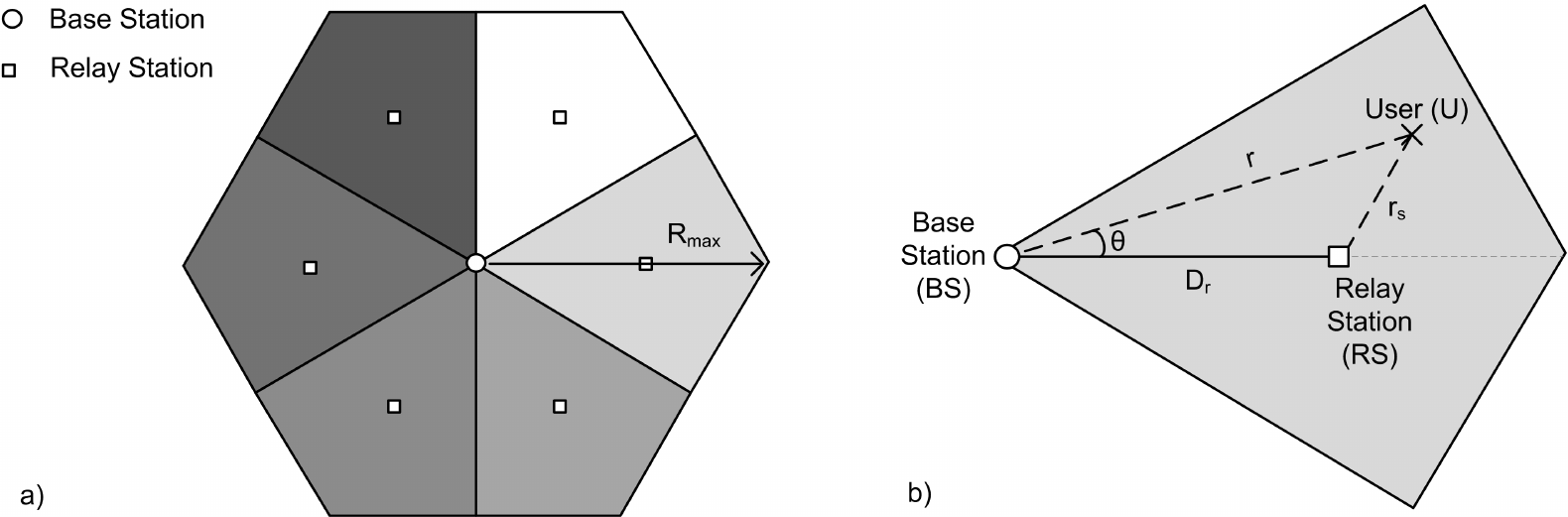}
\vspace*{10pt}
\caption{System model for a 6-sector urban cell}
\label{fig:cell_model}
\end{figure*}

\subsection{System model for relay placement analysis}

As presented in \cite{Peters2009} and depicted in Figure \ref{fig:cell_model}(a), we consider a 6-sector hexagonal cell. Note that, however, the analysis presented in this paper can be adapted to other cell models, such as circular shape or varying number of sectors. 
 $\mathsf{R}_{\text{cov}}$ refers to the maximum radius, which includes the coverage extension provided by the relay station. 
The base station is located at the center of the cell and each sector is provided with a relay station, located at a distance $\mathsf{D}_r$ from the base station.
The user is positioned at a distance $r \leq \mathsf{R}_{\text{cov}}$ from the base station, with an angle $\theta$, as illustrated in Figure \ref{fig:cell_model}(b). $r_s$ refers to the distance from the user to relay.
We consider a three-dimensional geometrical model. We analyze the impact of the relay height on the system performance. We denote $\mathsf{H}_{B}$, $\mathsf{H}_{R}$ and $\mathsf{H}_{U}$ the heights of the base station, the relay station and the user respectively. 

The multiple access strategy allows orthogonality between users, such that only one user is served for a given time and frequency resource.
We assume that the relay operates in the same frequency resource as the user it serves (\textit{in-band} relaying).

\subsection{Channel model for half-duplex relaying}

We consider half-duplex relaying performed in time division.
Nevertheless, the following analysis and results are valid with other multiplexing schemes. Without loss of generality, we describe the channel model for uplink transmissions.
Here, a transmission of unitary length is carried out in two phases of equal duration. We denote the transmitted codewords $X_{1}$ and $X_{2}$ for the user at each phase respectively, and $X_r$ for the relay. $Y_r$ and $Y_1$ respectively are the received signals at the relay and destination at the end of phase 1, $Y_2$ is the received signal at the destination at the end of phase 2. Thus, the half-duplex relay channel is written as follows:
\begin{equation}
\begin{split}
Y_r & = h_s X_1 + Z_r \, ; \quad
Y_1 = h_d X_1 + Z_1  \\ 
Y_2 &= h_d X_2 + h_r X_r + Z_2 
\end{split}
\end{equation}
where $Z_r$, $Z_1$ and $Z_2$ are independent additive white Gaussian Noises (AWGN) with equal variance $N$. We respectively denote $h_d$, $h_s$ and $h_r$ the channel from user to base station (direct link), from user to relay and from relay to base station.  For the channel coefficients $\vert h_{i} \vert^2 = \frac{1}{\gamma_{i}}$, we use the path-loss models as proposed in the WINNER II model \cite{winner}: 
\begin{equation}
\begin{split}
\gamma_{i}[dB] = & A_{i} \log_{10}(d) + B_{i} + C_{i} \log_{10} \left(\frac{f_c}{5} \right) 
 \\ &
 + D_{i} \log_{10}\left((\mathsf{H}_\text{Tx}-1)(\mathsf{H}_\text{Rx}-1)\right)
 \end{split}
\label{eq:pathloss}
\end{equation}
where $\mathsf{H}_\text{Tx}$ and $\mathsf{H}_\text{Rx}$ are the respective heights (in meters) of transmitter and receiver, $d$ is the distance (in meters) between them and $f_c$ the carrier frequency (in GHz). The parameters $A_{i}$, $B_{i}$, $C_{i}$ and $D_{i}$ are constants dependent on the global location of the transmitter and receiver (street level, rooftop...).
For convenience, we will use the following notation
\begin{align}
& \vert h_{i} \vert^2 = \frac{1}{\gamma_{i}} = \frac{1}{K_{i} d^{A_{i} /10}} \quad \text{with} 
\label{eq:Ki} \\ & 
K_{i} = 10^{B_{i}/10}\left(\frac{f_c}{5}\right)^{C_{i} /10} \left((\mathsf{H}_\text{Tx}-1)(\mathsf{H}_\text{Rx}-1)\right)^{D_{i}/10} .
\nonumber
\end{align}

Since the relay height is allowed to vary from below to above rooftop, the propagation environment for $h_s$ and $h_r$ changes notably. However, the WINNER II project does not provide a continuous path-loss model as a function of the relay height. We will therefore consider the two following situations: 
1/ \textit{Vicinity relay:} the relay station is located below rooftop, implying that the user-to-relay link is generally strong but at the cost of a weaker relay-to-base station link; 
2/ \textit{Base-station-like relay:} the relay station is located above the rooftop, like a base station, the $h_r$ link is very strong and can be assumes as free space (FS). Nevertheless, the user experiences the same path-loss model to access the relay as the base station.
We refer the reader to Appendix \ref{App:pathloss} for further details on the path-loss model. Note that this paper only considers path-loss, and not fading. The analysis and the proposed model can thus be understood as the mean coverage and relaying area over a sufficient period of time such that fading is averaged out.

In this channel model, we assume reasonably low inter-cell interference, as a starting point to create a baseline model. Indeed, techniques such as Fractional Frequency Reuse (FFR), scheduling based on inter-base-station coordination and directional antennas between the relay and the base station, have been spurred for next cellular OFDMA systems to suppress or avoid interference from neighbouring cells \cite{book_LTE_Fund}.

\section{Relaying schemes and Power Allocations for Energy Efficiency}
\label{sec:schemes}

We present the three reference coding schemes for half-duplex time-division relaying which are used for performance analysis. Once again, and without loss of generality, we describe the schemes for uplink transmissions.
We consider Gaussian signaling rather than practical modulations and coding.
Gaussian signaling is not only suitable for theoretical investigation (by providing upper bound on practical system performance), but is also appropriated for analysis of practical systems. Indeed, practical systems such as LTE are OFDM-based and data is transmitted over numerous parallel frequency channels such that the output is well approximated by a complex Gaussian distribution, as developed in \cite{book_LTE_Fund, OFDM_Gaussian}.

For each coding scheme, the power allocation minimizes the energy consumption while maintaining a given user rate $\mathcal{R}$.
The user and the relay have individual power constraints over the two transmission phases within the same bandwidth, respectively denoted as $P_U^{(\max)}$ and $P_R^{(\max)}$. The power constraint of the base station is set to $P_B^{(\max)}$. At each phase, each node allocates to each transmitted codeword a portion of its available power. We consider \textit{in-band} relaying, where the relay and the user operate in the same frequency band.

\subsection{Direct Transmission (DTx)}

As the first reference, we consider direct transmissions from user to base station. For fair comparison, direct transmission scheme is performed over the same duration as the relaying scheme, and not the half duration as generally done in the literature. By doing this, relaying does not create excess delay, nor uses more resource (time and/or frequency) than direct transmissions.
The user rate $\mathcal{R}$ is feasible as long as it is below the channel capacity, i.e.
\begin{align*}
\mathcal{R} \leq \log_2 \left( 1+ \frac{P_U^{(\max)} \vert h_d\vert^2}{N}\right) .
\end{align*}
The minimum required energy is equal to $\left( 2^\mathcal{R } - 1\right) \frac{N}{\vert h_d\vert^2}$. Outage occurs when the user power constraint $P_U^{(\max)}$ is not satisfied.

\subsection{Relay-aided Transmission (RTx)}

For relay-aided uplink transmissions, two schemes are considered: full-decode forward and partial decode-forward. For each, we use the power allocation that minimizes total energy consumption, i.e. consumption of the user and the relay station together. We furthermore assume that the user chooses to use either direct transmission (DTx) or relayed transmission (RTx) depending on which one consumes less energy.

\subsubsection{Full decode-forward scheme (Full-DF)}
\label{sec:two-hop}

During Phase 1, the user sends its message to the base station with rate $2\mathcal{R}$ and power $P_U$. Then, the relay decodes the message, re-encodes it with rate $2\mathcal{R}$ and forwards it to the destination during Phase 2 with power $P_R$.

In the literature, two main decoding techniques are generally considered
i) \textit{Two-hop relaying}, as used in practical system, in which the base station only considers the signal received from the relay and thus the relay is merely a repeater;
ii) \textit{Repetition-coded full decode-forward}, as defined in \cite{laneman2004}, in which the user repeats its message in Phase 2. Then, the base station uses MRC and combines the signals received from both the user and relay.
Rate bounds and energy consumption of both schemes can be similarly written using a parameter $\alpha$, with $\alpha = 0$ for two-hop relaying and $\alpha = 1$ for repetition-coded full decode-forward, as follows: 
\begin{align}
\left \lbrace
\begin{array}{rl}
\mathcal{R} &\leq \frac{1}{2} \log_2 \left(1+ \frac{P_U \vert {h}_s \vert^2}{N} \right) \\
\mathcal{R} &\leq \frac{1}{2} \log_2 \left(1+ \alpha \frac{P_U \vert {h}_d \vert^2}{N} +\frac{P_R \vert {h}_r \vert^2}{N} \right) 
\end{array}
\right. 
\end{align}
where the two lines describe the decoding constraints at the relay and the base station respectively. The power allocation set $(P_U,P_R)$ satisfies the following constraints:
\begin{align}
\frac{1}{2} P_U \leq & \; P_U^{(\max)}  
\quad \text{and} \quad
\frac{1}{2} P_R \leq \; P_R^{(\max)} \, .
\end{align}
The allocation set which minimizes the energy consumption is  \vspace*{-12pt}
\begin{equation}
\begin{split}
P_U &= \left(2^{2 \mathcal{R}} -1\right) \frac{N}{\vert h_s \vert^2 }
\quad \text{and} \\
P_R &= \left(2^{2 \mathcal{R}} - 1 \right)\left( 1- \alpha \frac{\vert h_d \vert^2}{\vert h_s \vert^2}\right)  \frac{N}{\vert h_r \vert^2 } .
\label{eq:allocation_FullDF}
\end{split}
\end{equation}

However, we can show that, in an urban cellular network considering path-loss only, combining the signals received from both the user and relay station as done in repetition-coded full decode-forward only brings little energy gain compared to two-hop relaying. 
This small gain is present only when the direct link is strong but, in this case, direct transmission is most of the time more energy-efficient than relaying.
Subsequently, we will consider two-hop relaying and repetition-coded full decode-forward as two variants of the same reference scheme, denoted as full decode-forward (Full-DF).

\subsubsection{Energy-optimized partial decode-forward scheme (EO-PDF)}

As the second relay-aided scheme, we consider the partial decode-forward scheme optimized for energy. This scheme was proposed in our previous work \cite{Journal1}, where it is referred as G-EE. In this paper, however, we will denote this scheme as EO-PDF, for Energy-Optimized Partial Decode-Forward, for the consistency of notation.
In the EO-PDF scheme, only part of the initial message is relayed, the rest being sent via the direct link. The scheme is based on rate splitting and superposition coding at transmitter, and joint decoding at receivers. It uses beamforming between the relay and the user in the second transmission phase. This scheme arguably requires complex implementation and fine synchronization, however, from the aspect of information theory, it is both energy- and rate-optimal as discussed in \cite{Journal1}, and thus, gives the theoretical upper-bound of the performance that can be achieved with decode-forward based relaying.

The optimal rate and power allocation of this coding scheme depends on the desired user rate and is composed of three sub-schemes as follows: \\
1) If the desired user rate is below a certain threshold $\mathcal{R}^\star$, \\
\hspace*{15pt}$\bullet$ In Phase 1, the user sends $m$ with power $P_U$ and the relay decodes as $\tilde{m}$. \\
\hspace*{15pt}$\bullet$ In Phase 2, the relay sends $\tilde{m}$ with power $P_R$ and the user sends $m$ with power $P_U$. \\
The allocation $\left(P_U,P_R\right)$ minimizes the total energy consumption of the user and the relay.
\\
2) If the desired user rate is above $\mathcal{R}^\star$, the source splits its message into two parts $m_r$ and $m_d$, of rates $\mathcal{R}^{(r)}$ and $\mathcal{R}^{(d)}$ respectively, with $\mathcal{R} = \mathcal{R}^{(d)}+\mathcal{R}^{(r)}$. \\
\hspace*{15pt}$\bullet$ In Phase 1, the source sends $m_r$ with power $P_U^{(1)}$. The relay decodes as $\tilde{m}_r$.\\
\hspace*{15pt}$\bullet$ In Phase 2, the relay sends $\tilde{m}_r$ with power $P_R$ and the source sends $(m_r,m_d)$ with power $(P_U^{(2)},P_U^{(3)})$, using superposition coding. The base station jointly decodes $m_r$ and $m_d$.
\\ The power and rate allocation $\left(\mathcal{R}^{(r)},\mathcal{R}^{(d)},P_U^{(1)},P_U^{(2)},\right.$ $\left.P_U^{(3)},P_R\right)$ minimizes the total energy consumption, with $\frac{1}{2} \left( P_U^{(1)} + P_U^{(2)} +P_U^{(3)}\right) \leq P_U^{(\max)}$.
\\
3) If this second sub-scheme is in outage due to very high user rate, the same coding scheme is used, i.e. rate splitting, superposition coding and joint decoding, but with other power allocation. Here, $\left(\mathcal{R}^{(r)},\mathcal{R}^{(d)},P_U^{(1)},P_U^{(2)},P_U^{(3)},P_R\right)$ minimizes the energy consumption of the relay only.

We refer the reader to \cite{Journal1} for the detailed power and rate allocation, as well as for the expression of  the threshold $\mathcal{R}^\star$.

\section{Characterization of efficient relay placement}
\label{sec:model}

In this section, we define the relaying efficiency, both in terms of energy consumption and coverage extension. To do so, we introduce the notion of Relay Efficiency Area (REA), and build our geometrical model based on this REA for both uplink and downlink transmissions. Note that we will consider Cartesian coordinates $(x,y)$ and polar coordinates $(r,\theta)$ for the user position.

\subsection{On the necessity to model the relaying efficiency}

\begin{figure}
	{\centering
	\subfigure[using Full decode-forward (Full-DF)]{
	    \includegraphics[width=0.82\columnwidth]{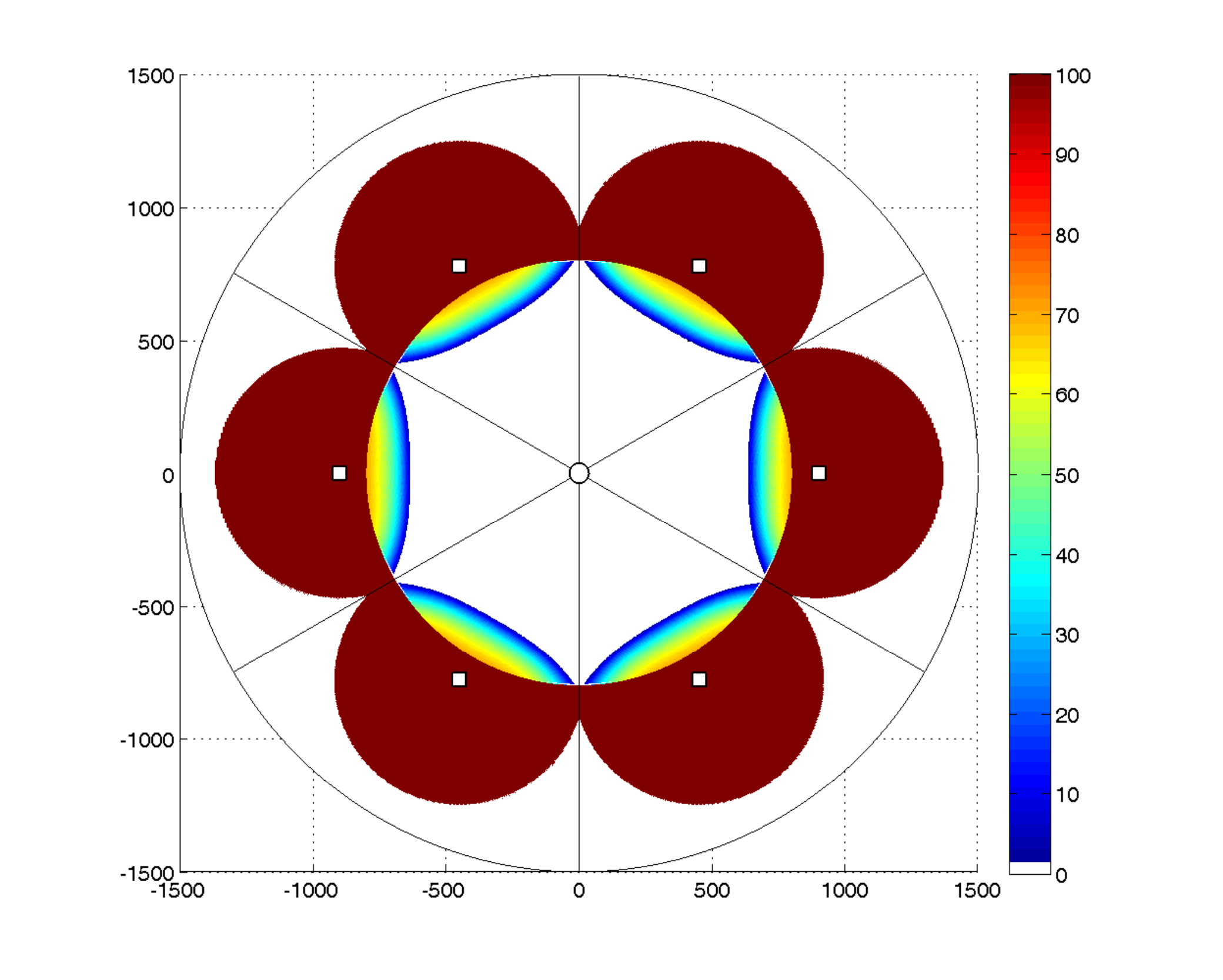}
	    \label{fig:Energy_gain_DF}
	} 
	\subfigure[performance upper-bound (EO-PDF)]{
	    \includegraphics[width=0.82\columnwidth]{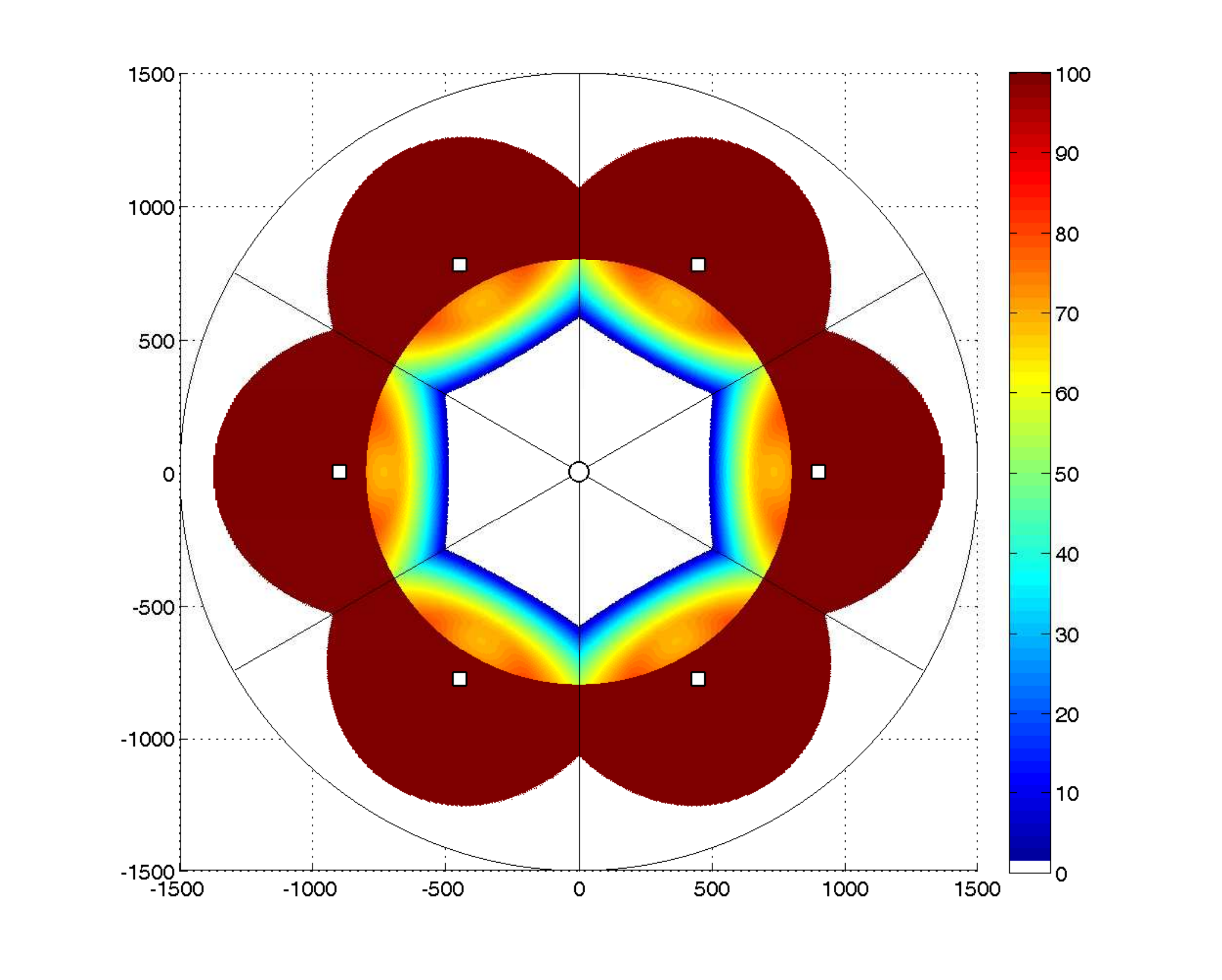}
	    \label{fig:Energy_gain_GEE}
	}
	 \caption{Energy gain compared to DTx ($\mathcal{R}=3$bit/s/Hz, vicinity relay, LOS conditions)}  
	\label{fig:Energy_gain} 
	}
\end{figure}

As a first insight of the potential energy gain brought by relaying, we plot in Figure \ref{fig:Energy_gain} the energy gain in percentage (i.e. $100\frac{E_{\text{DTx}}-E_{\text{RTx}}}{E_{\text{DTx}}}$) that is obtained by using either full decode-forward (Full-DF) or the energy-optimized partial decode-forward scheme (EO-PDF) over direct uplink transmission, as a function of the user location in the 6-sector cell. The gain is set to 100\% when direct transmission is not feasible, i.e. when relaying enhances the cell coverage.
As shown in the figure, relaying can significantly reduce the energy consumption.
Considering the cell-edge of conventional systems using direct transmission only,
an average of 46\% of energy gain is obtained with full decode-forward and 70\% with EO-PDF. In addition to energy savings, relaying also increases the radius of the corresponding hexagonal cell by 33\% for full decode-forward and by 56\% for EO-PDF.

Figure \ref{fig:Energy_gain} shows that conventional relaying schemes, such as two-hop relaying or repetition-coded full decode-forward, can provide notable performance enhancement, especially in terms of coverage extension. Nevertheless, those schemes are clearly suboptimal compared to the potential gains of decode-forward, upper-bounded by the EO-PDF scheme. It is then necessary to estimate the gains that implementing advanced coding schemes can bring over simpler schemes, both in terms of energy savings and coverage extension. Furthermore, the efficiency of the relay obviously depends on its location in the cell, as well as its propagation environment. 
Estimating the network performance cannot be reasonably done by simulating all possible relay network configurations and user locations. Thus, a model is needed to give insight of relay utilization and energy consumption.

\subsection{Analytical model for Relay Efficiency Area (REA)}

A relay-aided cell is characterized by the probability for a user to be served by the relay station and by the energy saved by using the relay. Those two performance criteria determine if a relay location is beneficial, and thus, characterize relay efficiency.
To investigate relay efficiency, we introduce here the concept of \textit{Relay Efficiency Area (REA)} for a cellular network served by relay stations. It is defined as the set of all user locations for which relaying is more energy-efficient than DTx or for which DTx is not feasible. Therefore, the REA covers both performance improvement and coverage extension.  

\begin{figure}
	{\centering
	\subfigure[Non-constrained cell]{
	    \includegraphics[width=0.35\textwidth]{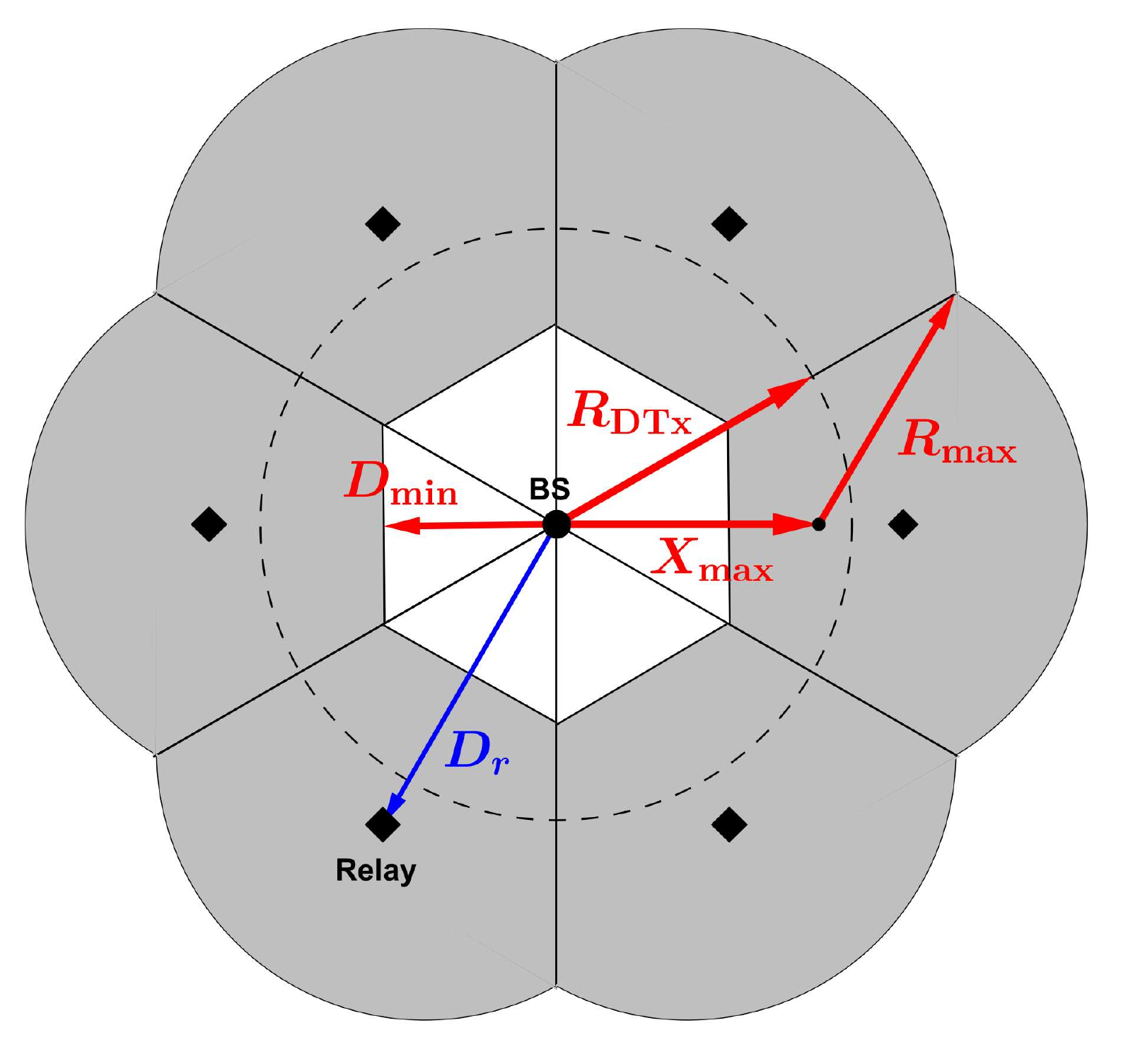}
	    \label{fig:REA_circle_model}
	}
	\subfigure[Hexagonal cell]{
	    \includegraphics[width=0.35\textwidth]{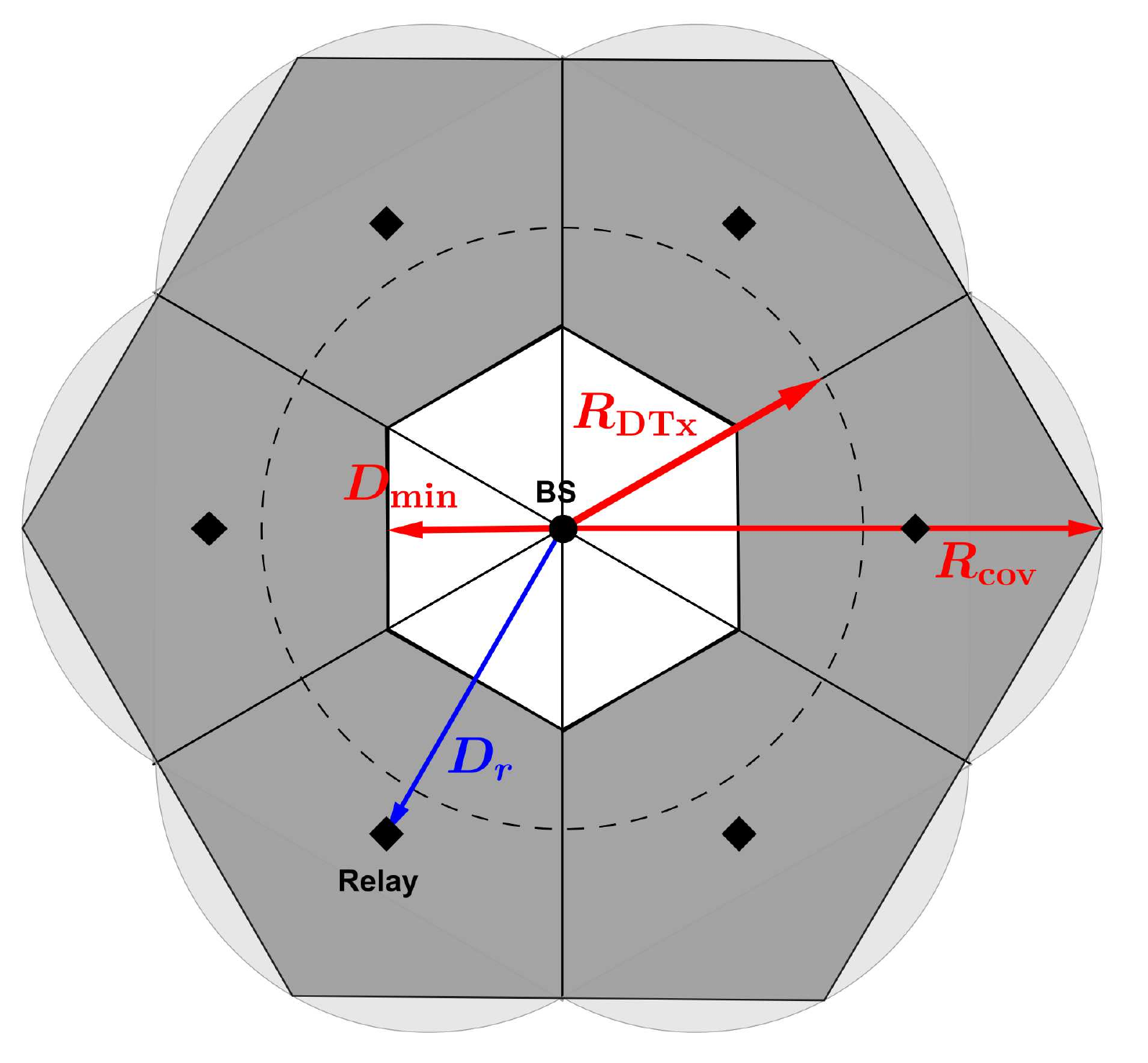}
	    \label{fig:REA_hexagone_model}
	}
	 \caption{A model for Relay Efficiency Area (REA)}  
	\label{fig:REA_model}
	}
\end{figure}

The simulated REA's of Full-DF and the EO-PDF scheme are illustrated in Figure \ref{fig:Energy_gain}, and its related model in Figure \ref{fig:REA_model}. In Figure \ref{fig:Energy_gain}, the REA corresponds to the cell area where the energy gain is strictly positive (non-white area).
When the user is close to the base station, the $h_d$ link is very strong and DTx consumes less energy than relaying. Otherwise, relaying is more energy-efficient or allows successful transmission while DTx is in outage.

We propose to characterize the REA of each coding scheme using four characteristic distances, as depicted in Figure \ref{fig:REA_circle_model}: $\mathsf{D}_{\min}$ and $\mathsf{R}_{\text{DTx}}$ for the inner bound, $\mathsf{X}_{\max}$ and $\mathsf{R}_{\max}$ for the outer bound. The inner bound is modelled by combining the strait line of equation  $x=\mathsf{D}_{\min}$ and a portion of the circle of radius $\mathsf{R}_{\text{DTx}}$ and centred at the base station, which refers to the minimum cell radius over which direct transmission is infeasible given outage requirement. The outer bound is characterized by $\mathsf{X}_{\max}$ and $\mathsf{R}_{\max}$. It is modelled by a portion of the circle of radius $\mathsf{R}_{\max}$ and centred at $(\mathsf{X}_{\max},0)$. The Relay Efficiency Area is thus modelled as follows:
\begin{equation}
\begin{split}
\text{REA} =  \left\{ (r, \theta) \; \in \; \right. &
\left( 
\left( \mathsf{D}_{\min} \leq  r \cos (\theta) \right) \; \cup \;
\mathcal{\bar{C}}\left(0, \mathsf{R}_{\text{DTx}}\right)
\right)
 \\  & \left.
\cap \; \left( \mathcal{C}\left(\mathsf{X}_{\max}, \mathsf{R}_{\max}\right) \right) \right \rbrace
\end{split}
\label{Eq:REA}
\end{equation}
where $(r, \theta)$ is the user location. $\mathcal{C}\left( \mathsf{X}, \mathsf{R}\right)$ stands for the portion of the disk centred at $\left(\mathsf{X}, 0\right)$ with radius $\mathsf{R}$ and delimited by the sector bounds.
$\mathcal{\bar{C}}$ stands for the exterior of this disk. Note that this model for REA is independent of the user distribution and of the cell shape. These factors are only considered for the computation of the performance metrics, e.g. the coverage radius, the probability of relaying and the average energy consumption.

Using this first characterization of REA, we derive the model for hexagonal cells and define the maximum cell radius $\mathsf{R}_{\text{cov}}$ that guarantees coverage for all users located inside the hexagonal cell without overlap, as depicted in Figure \ref{fig:REA_hexagone_model}. Three characteristic distances are sufficient to define this hexagonal model for REA: $\mathsf{D}_{\min}$, $\mathsf{R}_{\text{DTx}}$ and $\mathsf{R}_{\text{cov}}$, where
\begin{align}
\mathsf{R}_{\text{cov}} = \mathsf{X}_{\max} + \min \left \{
\mathsf{R}_{\max},
\sqrt{\frac{\left(2\mathsf{R}_{\max}\right)^2 - \left(\mathsf{X}_{\max}\right)^2}{3}}
\right \} .
 \label{eq:R_cov}
\end{align}
This expression of $\mathsf{R}_{\text{cov}}$ insures no coverage hole at sector-edge, contrary to the coverage radius defined in \cite{joshi2011,khakurel2012} . 
As we will show later, $\mathsf{R}_{\max}$ is function of the transmit power.
The area of a cell sector can then be expressed as follows.
\begin{align}
\mathcal{A}_{\text{sector}} = \frac{\sqrt{3}}{4} \mathsf{R}_{\text{cov}}^2 .
\label{eq:aire_}
\end{align}

These characteristic distances, as defined above, allow the direct computation of the overall probability of relaying a communication, given that the user is randomly located in the cell, as well as the average consumed energy to successfully transmit data.
In the following sections, we apply this model to compute these characteristic distances for each coding scheme and investigate the probability of relaying in the cell and the total average energy consumption.

\section{Analysis of Relay Efficiency Area for decode-forward schemes}
\label{sec:analysis_model}

We now compute the four characteristic distances $\mathsf{D}_{\min}$, $\mathsf{R}_{\text{DTx}}$, $\mathsf{X}_{\max}$ and $\mathsf{R}_{\max}$ for Full-DF and EO-PDF, as defined in Section \ref{sec:schemes}, and for both uplink and downlink transmissions.
These characteristic distances are derived from the channel conditions for transmission feasibility and for energy efficiency.

\subsection{Common conditions for relaying}

First, we focus on $\mathsf{R}_{\text{DTx}}$, the maximum radius over which DTx is infeasible. We have
\begin{align}
\mathsf{R}_{\text{DTx}} = \left(\frac{P_i}{K_d N \left(2^{\mathcal{R}}-1\right)} \right)^{10/A_d}
\end{align}
where $A_d$ and $K_d$ are given by the path-loss model and defined in Eq. \eqref{eq:pathloss} and \eqref{eq:Ki}, and where $P_i=P_U^{(\max)}$ for uplink and $P_i=P_{B}^{(\max)}$ for downlink.

Second, we define $\mathsf{D}_{\text{RTx}}$ for the uplink. It refers to the relaying condition $\vert h_d \vert^2 \leq \vert h_s \vert^2$ for which relaying can outperform DTx. At the limit, we get
\begin{align}
& \vert h_d \vert^2 = \vert h_s \vert^2 \label{eq:relaying_condition}
\\  \nonumber 
 \Leftrightarrow & \quad
\left(\frac{K_d}{K_s}\right)^{\frac{20}{A_s}} r ^{2\frac{A_d}{A_s}} - r^2 +2 \mathsf{D}_r \cos\left(\theta \right) r - \mathsf{D}_r^2 =0
\end{align}
where $\mathsf{D}_r$ is the distance between the relay and base station.
Given the cell sector bounds and regarding Cartesian coordinates, we can show that the derivative of x with respect to y is small, such that Eq. \eqref{eq:relaying_condition} can be approximated by a straight line. Thus, in the proposed model, we approximate the relaying condition $\vert h_d \vert^2 \leq \vert h_s \vert^2$ by $x \geq \mathsf{D}_{\text{RTx}}$, where
$\mathsf{D}_{\text{RTx}}$ is defined as the smallest positive solution of $r$ in Eq. \eqref{eq:relaying_condition} with $\theta = 0$, i.e.
\vspace*{-10pt}
\begin{align}
\left(\frac{K_d}{K_s}\right)^{\frac{20}{A_s}} r ^{2\frac{A_d}{A_s}} = \left(r - \mathsf{D}_r\right)^2
\end{align}
In particular, when $h_s$ and $h_d$ have the same path-loss exponent, as it is the case when both $h_s$ and $h_d$ are in LOS, we get:
\vspace*{-10pt}
\begin{align}
\mathsf{D}_{\text{RTx}} = \frac{\mathsf{D}_r}{\left(\frac{K_d}{K_s} \right)^{\frac{10}{A_s}} +1} .
\end{align}
This relaying condition must be satisfied for both Full-DF and EO-PDF. In practice however, this condition is not limiting for Full-DF since, for this coding scheme, transmission is over two phases of half-duration and data is thus sent at twice the desired rate. Actually, the user decides if relay-aid is necessary based on energy consumption, as we will see in next paragraph. On the contrary, the relaying condition $\vert h_d \vert^2 \leq \vert h_s \vert^2$ is relevant for the EO-PDF scheme since only part of the message is relayed, leading to sufficient energy gain to compensate for the shorter transmission duration.

With regards to downlink, this condition becomes $\vert h_d \vert^2 \leq \vert h_r \vert^2$, which is always satisfied since the relay-to-base-station link is very strong.

\subsection{Conditions for energy efficiency and outage of full decode-forward (Full-DF)}

\subsubsection{Energy-efficiency of full decode-forward}
\label{sec:EE_Full_DF}

First, we focus on energy efficiency and we derive the conditions for which the energy consumption using DTx (denoted $E_{\text{DTx}}$) is greater than the consumption using full decode-forward (denoted $E_{\text{DF}}$). This energy-efficient condition can be seen as the dual of the throughput-oriented condition $\log_2 \left(1 + \frac{P_s \vert h_d \vert^2}{N}\right) \leq \frac{1}{2}\log_2 \left(1 + \frac{2 P_s \vert h_s \vert^2}{N}\right)$.  By using Eq. \eqref{eq:allocation_FullDF}, we get
\begin{align}
E_{\text{DTx}} \geq & E_{\text{DF}} \label{eq:RDF1}
\\ \nonumber \Leftrightarrow \quad
\left( 2^{\mathcal{R}}-1 \right) \frac{N}{\vert h_d \vert^2} \geq &
\left( 2^{2\mathcal{R}}-1 \right) \frac{N}{2\vert h_s \vert^2} 
\\ \nonumber &
+ \left( 2^{2\mathcal{R}}-1 \right) \frac{N}{2\vert h_r \vert^2} \left(1 - \alpha \frac{\vert h_d \vert^2}{\vert h_s \vert^2} \right)
\\ \nonumber \Leftrightarrow \quad 
\alpha K_s K_r \mathsf{D}_r^{A_r/10} \leq &
K_d r^{A_d/10} \left[ K_s - \left(\frac{2}{2^{\mathcal{R}}+1} K_d r^{A_d/10} 
\right. \right. \\ \nonumber  & \left. \left.
- K_r \mathsf{D}_r^{A_r/10} \right)\frac{1}{r_s^{A_s/10}}\right] 
\\ \text{where } \quad r_s^2  = \mathsf{D}_r^2 & + r^2 - 2 \mathsf{D}_r r \cos(\theta) . \nonumber
\end{align}

Note that Eq. \eqref{eq:RDF1} is symmetric in $h_s$ and $h_r$, meaning that it is valid for both uplink and downlink.
We can also show that a solution of Eq. \eqref{eq:RDF1} for $\alpha = 0$ is almost solution for $\alpha = 1$.
This implies that the bounds for two-hop relaying can approximate the ones of repetition-coded full DF. Simulations in Section \ref{sec:simulation} show that the REA of repetition-coded full DF is only increased by less than 1\% compared to two-hop relaying.

The solution of Eq. \eqref{eq:RDF1} can be well approximated by the intersection of the half-plane defined by $x \geq \mathsf{D}_{\text{DF}}^{(e)}$ (inner bound) with the disk $\mathcal{C}\left( \mathsf{X}_{\text{DF}}^{(e)} , \mathsf{R}_{\text{DF}}^{(e)} \right)$ (outer bound), where superscript $^{(e)}$ stands for energy efficiency.
 $\mathsf{X}_{\text{DF}}^{(e)}$ and $\mathsf{R}_{\text{DF}}^{(e)}$ can be computed easily as follows. Noting that there exists only one circle with its center on the x-axis and going through any two given points of the plane, using geometrical principles, it is therefore sufficient to find the two user locations for which $E_{\text{DTx}} = E_{\text{DF}}$ with $\theta = 0$ and $\theta = \frac{\pi}{6}$ to deduce $\mathsf{X}_{\text{DF}}^{(e)}$ and $\mathsf{R}_{\text{DF}}^{(e)}$.
We propose to reject all relay positions for which $\mathsf{R}_{\text{cov}} < \mathsf{R}_{\text{DTx}}$, i.e. for which $\mathcal{C}\left( \mathsf{X}_{\text{DF}}^{(e)} , \mathsf{R}_{\text{DF}}^{(e)} \right)$ is strictly included in $\mathcal{C}\left( 0 , \mathsf{R}_{\text{DTx}}\right)$, i.e. the area covered by the base station alone.
Consequently, $\mathcal{C}\left( \mathsf{X}_{\text{DF}}^{(e)} , \mathsf{R}_{\text{DF}}^{(e)} \right)$ is only used to determine the acceptance or rejection of a relay configuration, but does not appear in the expression of the REA itself, as defined in Eq. \eqref{Eq:REA}.

\subsubsection{Outage of full decode-forward}

Next, we focus on the outage condition. Assuming the relay power is sufficient, outage occurs when the transmission between the relay and user is not feasible, i.e. when the channel $h_s$ is too weak, given the power constraint and user rate. Note that direct transmission is not feasible either in this case. Relayed transmission is in outage for all users positions outside the disk centred at the relay station and of radius:
\begin{align}
\mathsf{R}_{\text{DF}}^{(o)} = \left(\frac{2 P_i}{K_s N \left( 2^{2\mathcal{R}}-1 \right)} \right)^{\frac{10}{A_s}} . \label{Eq:R_DF_o}
\end{align}
where $^{(o)}$ stands for outage and where $P_i=P_U^{(\max)}$ for uplink and $P_i=P_{R}^{(\max)}$ for downlink.

\subsubsection{Characteristic distances for full decode-forward}

Based on the previous analysis, we can deduce the characteristic distances for full decode-forward as follows:
\begin{align}
\hspace*{-10pt}\left\lbrace \begin{array}{l l}
\mathsf{D}_{\min}^{(\text{DF})} & =\max \left\{ \mathsf{D}_{\text{RTx}}, \mathsf{D}_{\text{DF}}^{(e)} \right \} \quad \text{with } \quad \mathsf{D}_{\text{DF}}^{(e)} \leq \mathsf{R}_{\text{DTx}}\\
\mathsf{X}_{\max}^{(\text{DF})} & = \mathsf{D}_{r} \; ; \quad
\mathsf{R}_{\max}^{(\text{DF})} = \mathsf{R}_{\text{DF}}^{(o)} .
\end{array}
\right. 
\end{align}
where $\mathsf{D}_{\text{RTx}}$ is given by Eq. \eqref{eq:relaying_condition}, $\mathsf{D}_{\text{DF}}^{(e)}$ by Eq. \eqref{eq:RDF1} and $\mathsf{R}_{\text{DF}}^{(o)}$ by Eq. \eqref{Eq:R_DF_o}.
These values define the Relay Efficiency Area of full decode-forward $\text{REA}^{(\text{DF})}$, as defined in Eq. \eqref{Eq:REA}.
Note that REA of Full-DF regarding uplink is strictly included in the REA for downlink. $\mathsf{D}_{\min}^{(\text{DF})}$, which indicates whether the user should use DTx or RTx is the same, but the coverage radius is extended for downlink since the relay station disposes of more power than the user.

\newcounter{MYtempeqncnt}
\begin{figure*}[!t]
\setcounter{MYtempeqncnt}{\value{equation}}
\setcounter{equation}{18}
\small{
\begin{align}
&\left \lbrace
\begin{array}{ll}
\varphi = \phi = \frac{\pi}{6}
&
\text{if } \mathsf{D}_{\min} \leq \min \left(\frac{\sqrt{3}}{2}\mathsf{R}_{\text{DTx}}, \frac{3}{4}\mathsf{R}_{\text{cov}}\right)
\\
\varphi = \arccos \left(\frac{\mathsf{D}_{\min}}{\mathsf{R}_{\text{DTx}}}\right); \phi = \frac{\pi}{6}
&
\text{if } \frac{\sqrt{3}}{2}\mathsf{R}_{\text{DTx}} \leq  \frac{3}{4}\mathsf{R}_{\text{cov}}
 \quad \text{and } \frac{\sqrt{3}}{2}\mathsf{R}_{\text{DTx}} \leq \mathsf{D}_{\min} 
\\
\varphi = \phi = \arctan \left(\sqrt{3} \left( \frac{\mathsf{R}_{\text{cov}} }{\mathsf{D}_{\min}} -1\right)\right)
&
\text{if } \frac{3}{4}\mathsf{R}_{\text{cov}} \leq  \frac{\sqrt{3}}{2}\mathsf{R}_{\text{DTx}} 
 \quad \text{and } \frac{3}{4}\mathsf{R}_{\text{cov}} \leq \mathsf{D}_{\min} \leq \mathsf{X}
%
\\
\varphi = \arccos \left(\frac{\mathsf{D}_{\min}}{\mathsf{R}_{\text{DTx}}}\right); \phi = \frac{\pi}{6} - \arccos \left(\frac{\sqrt{3}\mathsf{R}_{\text{cov}}}{2\mathsf{R}_{\text{DTx}}}\right)
&
\text{if } \frac{3}{4}\mathsf{R}_{\text{cov}} \leq  \frac{\sqrt{3}}{2}\mathsf{R}_{\text{DTx}} 
 \quad \text{and } \frac{3}{4}\mathsf{R}_{\text{cov}} \leq  \mathsf{X} \leq \mathsf{D}_{\min}
\end{array}
\right.
\label{eq:angles}
\end{align}
}
\hrulefill
\setcounter{equation}{\value{MYtempeqncnt}}
\end{figure*}

\subsection{Conditions for energy efficiency and outage of the EO-PDF scheme}

The computation of the characteristic distances for the EO-PDF scheme follows the same basis as for full decode-forward.

\subsubsection{Energy-efficiency of EO-PDF}

We recall that the inner-bound is modelled by the combination of the strait line of equation $x=\mathsf{D}_{\min}$ and a portion of the circle centred at the base station and of radius $\mathsf{R}_{\text{DTx}}$. $\mathsf{D}_{\min}$ usually depends on both the relaying condition defined in Eq. \eqref{eq:relaying_condition} and on the energy-efficiency condition $E_{\text{DTx}} \geq E_{\text{EO}}$.

With regards to uplink, simulations show that there can exist user locations for which direct transmissions is more energy-efficient than the EO-PDF scheme. Indeed, in DTx, messages are sent once over the two phases of transmission, with rate $\mathcal{R}$, whereas in the EO-PDF scheme, messages are sent twice, over each phase, with rate $2\mathcal{R}$. For some specific user locations, the energy gain obtained by EO-PDF does not compensate the loss due to transmitting at a rate of $2 \mathcal{R}$ during each phase instead of $\mathcal{R}$ during both phases, and thus, direct transmission becomes more energy-efficient.
Nevertheless, simulations show that these user locations are negligible for relay-aided cellular networks. First, they occur only for uncommon relay configurations, specifically when the relay is very close to base station and with very low user rates. Moreover, the area of concern remains small compared to the cell size and the energy gain obtained by DTx over EO-PDF is admittedly positive but almost zero. For these reasons, we can consider that $\mathsf{D}_{\min}$ reduces to the relaying condition $\vert h_d \vert ^2 \leq \vert h_s \vert^2$, as described in Eq. \eqref{eq:relaying_condition}, and is not limited by the energy-efficiency constraint as it is the case for full decode-forward.

Considering downlink transmission, and as said earlier, the relaying condition $\vert h_d \vert^2 \leq \vert h_r \vert^2$ is always satisfied and $\mathsf{D}_{\text{RTx}}$ for downlink is large compared to the cell radius. The inner-bound $\mathsf{D}_{\min}$ thus only depends on $\mathsf{D}_{\text{EO}}^{(e)}$, given by the energy-efficiency condition $E_{\text{DTx}} \geq E_{\text{EO}}$.  $\mathsf{D}_{\text{EO}}^{(e)}$ is computed numerically as the user-to-base-station distance for which $E_{\text{DTx}} = E_{\text{EO}}$ with $\theta=0$.

\subsubsection{Outage of EO-PDF}

The outer bound for the EO-PDF scheme is modelled by a portion of circle centred at $(\mathsf{X}_{\text{EO}}^{(o)},0)$ and of radius $\mathsf{R}_{\text{EO}}^{(o)}$. Both are determined by the outage condition and are computed similarly to $\mathsf{X}_{\text{DF}}^{(e)}$ and $\mathsf{R}_{\text{DF}}^{(e)}$ for full decode-forward in Section \ref{sec:EE_Full_DF}, using the coordinates of two user locations at the outage limit. However, generally, $\mathsf{X}_{\text{EO}}^{(o)} \neq \mathsf{D}_r$ and $\mathsf{R}_{\text{EO}}^{(o)} \neq \mathsf{R}_{\text{DF}}^{(o)}$.

\subsubsection{Characteristic distances for the EO-PDF scheme}

Based on the previous analysis, we can deduce the characteristic distances for the EO-PDF scheme as follows:
\begin{align}
\left\lbrace \begin{array}{l}
\mathsf{D}_{\min}^{(\text{EO})} =\max \left\{ \mathsf{D}_{\text{RTx}}, \mathsf{D}_{\text{EO}}^{(e)} \right \} \; ; \\
\mathsf{X}_{\max}^{(\text{EO})} = \mathsf{X}_{\text{EO}}^{(o)} \; ; \quad
\mathsf{R}_{\max}^{(\text{EO})} = \mathsf{R}_{\text{EO}}^{(o)} .
\end{array}
\right. 
\end{align}
where $\mathsf{D}_{\text{RTx}}$ is given by Eq. \eqref{eq:relaying_condition}. These values define the Relay Efficiency Area of the EO-PDF scheme $\text{REA}^{(\text{EO})}$, as defined in Eq. \eqref{Eq:REA}.

Note that for both Full-DF and EO-PDF, we have
$\mathsf{D}_{\min} \leq \mathsf{D}_r$. This shows that models based on inner and outer regions as in \cite{Chandwani2010}
do not reflect the performance gains obtained by relaying and that the cell area for energy-efficient relaying is much wider than the classical outer region.

\section{Estimated probability of relaying and average energy consumption}
\label{sec:P_RTx_energy}

So far, we have introduced the Relay Efficiency Area and derived characteristic distances for both Full-DF and EO-PDF. Based on this model, we now derive the estimated probability of relaying and average energy consumption of the cell. 

\subsection{Probability of Relaying}

Here, we analyze the probability of relaying a communication. We assume that the user is randomly located in the cell and that the location distribution is uniform. Such distribution allows us to derive a simple expression of the probability of relaying.

The probability of relaying, denoted $\mathbb{P}_{\text{RTx}}$, is obtained by the ratio of relaying area vs. total area and is expressed as follows for hexagonal cells.
\begin{align}
\mathbb{P}_{\text{RTx}} = &
1- \frac{8}{\sqrt{3} \mathsf{R}_{\text{cov}}^2}
 \left(\frac{\mathsf{D}_{\min}^2}{2} \tan \left( \varphi \right) 
 +\frac{\mathsf{R}_{\text{DTx}}^2}{2} \left(\phi - \varphi \right) 
  \right. \nonumber
 \\ & \left.
 +\frac{3}{8} \mathsf{R}_{\text{cov}}^2 \tan \left( \frac{\pi}{6} - \phi \right)  \right)
\label{Eq:Prob_RTx}
\end{align}
where we assume $\mathsf{R}_{\text{DTx}} \leq \mathsf{R}_{\text{cov}}$. $\varphi$ and $\phi$ are as in Eq. \eqref{eq:angles} at the top of the page, with $\mathsf{X} = \frac{3}{4} \mathsf{R}_{\text{cov}} + \frac{1}{4} \sqrt{4\mathsf{R}_{\text{DTx}}^2 - 3 \mathsf{R}_{\text{cov}}^2}.$
\addtocounter{equation}{1}

Conditions on $\phi$ and $\varphi$ are defined based on geometrical principles of a cell sector and depend on whether $\mathsf{D}_{\min} \leq  x$ or $(x,y) \in \mathcal{\bar{C}}\left(0, \mathsf{R}_{\text{DTx}}\right)$, as defined in \eqref{Eq:REA}, which is the limiting condition for the REA. The complete proof can be found in Appendix B. 
Note that, for circular shape cells, we have $\phi = \frac{\pi}{6}$ and the constraints regarding $\frac{3}{4}\mathsf{R}_{\text{cov}}$ are relaxed.

\subsection{Average energy consumption per user transmission}

We now focus on the energy consumption, averaged over all user locations. 
The expressions for downlink and uplink are similar, and we thus only focus on the uplink case.
As before, we assume that the user is randomly located in the cell and that its distribution is uniform.
We respectively denote $\mathbb{E} \left[E_{\text{DTx}}\right]$, $\mathbb{E} \left[E_{U}\right]$ and $\mathbb{E} \left[E_{R}\right]$ the average energy consumed by the user when DTx is used, by the user when RTx is used and by the relay station. The total average energy consumption, referred as $\mathbb{E} \left[E\right]$, can thus be computed as follows. 
\begin{align}
\hspace*{-10pt}\mathbb{E} \left[E\right] = 
\left(1-\mathbb{P}_{\text{RTx}}\right) \mathbb{E} \left[E_{\text{DTx}}\right]
+\mathbb{P}_{\text{RTx}} \left( \mathbb{E} \left[E_{U}\right] + \mathbb{E} \left[E_{R}\right]\right).
\label{eq:energy}
\end{align}
This expression reflects both the energy consumed by the user only when DTx is used, and the energy consumed by both the user and the relay station when RTx is employed.

We first obtain $\mathbb{E} \left[E_{\text{DTx}}\right]$, the average energy consumed by a user who is not located in the REA. It depends on $\mathsf{D}_{\min}$, $\mathsf{R}_{\text{DTx}}$ and $\mathsf{R}_{\text{cov}}$ as follows:
\begin{align}
& \mathbb{E} \left[E_{\text{DTx}}\right]  \nonumber \\
= & \underset{(x,y) \in \overline{\text{REA}}}{\iint} \left(2^{\mathcal{R}}-1\right) \frac{N}{\vert h_d \vert^2} \mathbb{P}\left((x,y) \in \overline{\text{REA}}\right) dx dy \label{eq:ave_energy_dtx}
 \\
= & \frac{2 \left(2^{\mathcal{R}}-1\right)N  K_d } {\mathcal{A}_{\text{sector}} \left(1-\mathbb{P}_{\text{RTx}}\right)}
\left(\frac{A_d}{10} + 2\right)
\nonumber \\ &
 \times \left[
\mathsf{D}_{\min}^{A_d/10+2} 
\int_0^{\varphi}\left(\frac{1}{\cos(\theta)}\right)^{A_d/10+2} d\theta 
 + 
 \left(\mathsf{R}_{\text{DTx}} \right)^{A_d/10+2} \left(\phi - \varphi\right)
 \right. \nonumber \\ 
& \left.
 +
\left(\sqrt{3}\mathsf{R}_{\text{cov}} \right)^{A_d/10+2} 
\int_{\phi}^{\frac{\pi}{6}}\left(\frac{1}{\sin(\theta)+\sqrt{3}\cos(\theta)}\right)^{A_d/10+2} d\theta \right] 
\nonumber 
\end{align}
where $\varphi$ and $\phi$ are as defined in Eq. \eqref{eq:angles}.

Second, the average energy consumed by the relay depends on the coding scheme. Although this energy cannot be derived in closed-form for the EO-PDF scheme, we provide in next paragraph the expression of the average energy for the full decode-forward scheme.

\paragraph*{Average energy consumed by the Full-DF scheme}

We analyze the average total energy consumption when the user is located in the REA, using full decode-forward. Recall that, as exemplified by simulations, repetition-coded full decode-forward only brings a few percent of energy savings compared to two-hop relaying. Therefore, to compute the average energy consumption, we focus on the two-hop relaying scheme, as defined in Section \ref{sec:two-hop}.
The average energy is denoted as $\mathbb{E} \left[E_{U}^{(\text{DF})}\right]$ for the user consumption and $\mathbb{E} \left[E_{R}^{(\text{DF})}\right]$ for the relay consumption. Both depend on $\mathsf{D}_{\min}$, $\mathsf{R}_{\text{DTx}}$ and $\mathsf{R}_{\text{cov}}$.

Similarly to $\mathbb{E} \left[E_{\text{DTx}}\right]$ in Eq. \eqref{eq:ave_energy_dtx}, we have 
\begin{align}
& \mathbb{E} \left[E_{U}^{(\text{DF})}\right] \nonumber 
\\
= & \underset{(x,y) \in \text{REA}}{\iint} \left(2^{2\mathcal{R}}-1\right) \frac{N}{2 \vert h_s \vert^2} \mathbb{P}\left((x,y) \in \text{REA}\right)  dx dy  \\
& = \frac{\left(2^{2\mathcal{R}}-1\right)N K_s } {\mathcal{A}_{\text{sector}} \mathbb{P}_{\text{RTx}}}
\left[ \int_0^{\varphi} \int_{\frac{ D_{\min}}{\cos(\theta)}} ^{r_{\max}\left( \theta \right)}  r_s^{A_s/10} r dr d\theta 
 \nonumber \right. 
\\
& \quad  \left. 
\nonumber 
+ \int_\varphi^{\phi} \int_{\mathsf{R}_\text{DTx}}^{r_{\max}\left( \theta \right)} r_s^{A_s/10} r dr d\theta
\right]
\nonumber \\
& \text{where} \quad
\left\lbrace
\begin{array}{ll}
r_s &= \sqrt{\mathsf{D}_r^2 + r^2 - r\mathsf{D}_r \cos (\theta)} \\
r_{\max}\left( \theta \right) & =\frac{\sqrt{3}\mathsf{R}_{\text{cov}}}{\sin(\theta)+\sqrt{3}\cos(\theta)}
\end{array}
\right. \nonumber
\end{align}
In this, $r_{\max}\left( \theta \right) $ describes the cell bound. $\varphi$ and $\phi$ are as defined in Eq. \eqref{eq:angles}.
Note that, in most of cases, we have $A_s=40$, for which the integral $\int r_s^{A_s/10} r dr$ becomes a polynomial and is equal to $\int \left( \mathsf{D}_r^2 + r^2 - \mathsf{D}_r \cos (\theta) r  \right)^2 r dr $, which can be computed easily.

We compute the average relay energy consumption $\mathbb{E} \left[E_{R}^{(\text{DF})}\right]$ as follows. 
\begin{align}
\mathbb{E} \left[E_{R}^{(\text{DF})}\right] 
&= \underset{(x,y) \in \text{REA}}{\iint} \left(2^{2\mathcal{R}}-1\right) \frac{N}{2 \vert h_r \vert^2} \mathbb{P}\left((x,y) \in \text{REA}\right)  dx dy \nonumber \\
& = \left(2^{2\mathcal{R}}-1\right) \frac{N}{2} K_r \mathsf{D}_r ^{A_r/10} 
\end{align}
where $\mathcal{A}_{\text{sector}}$ and $\mathbb{P}_{\text{RTx}}$ are as defined in Eq. \eqref{eq:aire_} and Eq. \eqref{Eq:Prob_RTx} respectively.
These expressions for the user and relay consumptions conclude the analysis of Relay Efficiency Area.

\section{Simulation results and energy-efficient relay location}
\label{sec:simulation}

In this section, we simulate the performance obtained by relay-aided transmissions and highlight energy-efficient relay configurations.
For simulation, if not specified, we use the parameters of Table \ref{sim_param}. 
Also note that we do not consider configurations where the direct link $h_d$ is LOS but the user-to-relay link $h_s$ is NLOS. In this case, DTx is performed due to the bad quality of the relaying path. We also assume that the relay remains inside the cell, i.e. $\mathsf{D}_r \leq \mathsf{R}_{\text{cov}}$.

\begin{table}
\centering \small{ \begin{tabular}{|c|l|} 
\hline
Carrier frequency & $f_c$=2.6GHz   \\
Noise & $N=5.10^{-13}$ \\
Max power & $P_U^{(\max)}$=500mW \\
& $P_R^{(\max)}$=1W  \\
Node heights & $\mathsf{H}_{B}$=30m, $\mathsf{H}_{U}$=1.5m \\
 & 10m $\leq \mathsf{H}_{R} \leq $ 30m \\
 User rate & $\mathcal{R}$= 3bits/s/Hz \\
 Other & $h_d$ / $h_s$ / $h_r$ LOS \\
\hline
\end{tabular}}
\caption{Default simulation parameters}  
\label{sim_param}
\end{table}

\subsection{Validation of the model}

\begin{figure}

	    \centering \hspace*{-30pt} \includegraphics[width=0.87\columnwidth]{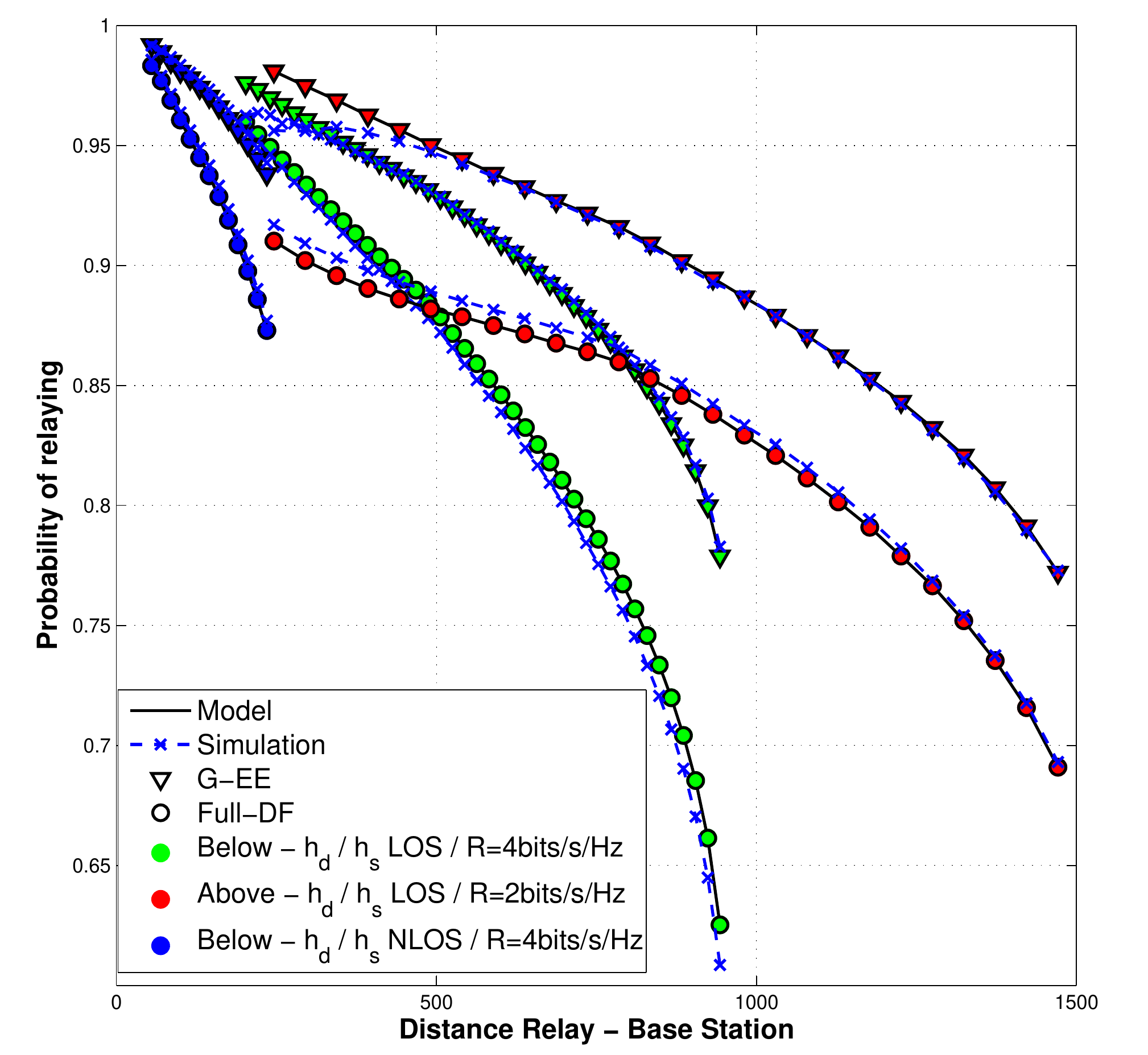}

	\caption{Validation of the probability of relaying $\mathbb{P}_{\text{RTx}}$}  
	 \label{fig:validity_model_prob}

\end{figure}

We first validate the REA model, i.e. the proposed characteristic distances $\mathsf{R}_{\text{DTx}}$, $\mathsf{R}_{\text{cov}}$ and $\mathsf{D}_{\min}$. An error occurs for a given user position if relaying is decided according to the REA model while DTx would be in reality more energy-efficient, or reversely, if DTx is decided rather than relaying. 
The validity of the proposed model mostly depends on the approximation of $\mathsf{D}_{\min}$, which embraces the relaying condition, as written in Eq. \eqref{eq:relaying_condition}, and the energy-efficiency condition, i.e. $E_{\text{DTx}} \geq E_{\text{RTx}}$.
We plot in Figure \ref{fig:validity_model_prob} the probability of relaying as a function of the relay to base station distance $\mathsf{D}_r$. This probability is evaluated both by simulation and computation of Eq. \eqref{Eq:Prob_RTx}.
We consider both relay coding schemes in various propagation environments. We see that the proposed characteristic distances lead to a probability of using the relay within 3\% of the simulated probability. Thus, wrong decision occurs only for 3\% of all the possible user positions in the cell, those errors being mostly located around the strait line $x=\mathsf{D}_{\min}$.
Note that the REA model approaches the simulated consumption within 1\%. Indeed, in the neighbourhood of $x=\mathsf{D}_{\min}$ where most of erroneous decisions occurs, distances from user to relay and user to base station do not vary significantly enough to imply notable difference in the energy consumption for either RTx or DTx. 

\subsection{Performance analysis on coverage extension and energy efficiency using Full-DF}

In this paragraph, we analyze separately the coverage extension allowed by relaying given power constraints, and the minimal energy consumed for a given cell coverage.
We first base our analysis on uplink transmissions for Full-DF, since this scheme is widely considered for relaying in future cellular standard such as LTE. As performance criteria, we consider the coverage and energy gains when the relay is optimally located, and analyze how much performance is degraded when the relay is away from this optimal location.

\begin{figure}
	{\centering 
	\subfigure[Maximal coverage extension]{
	    \includegraphics[width=0.87\columnwidth]{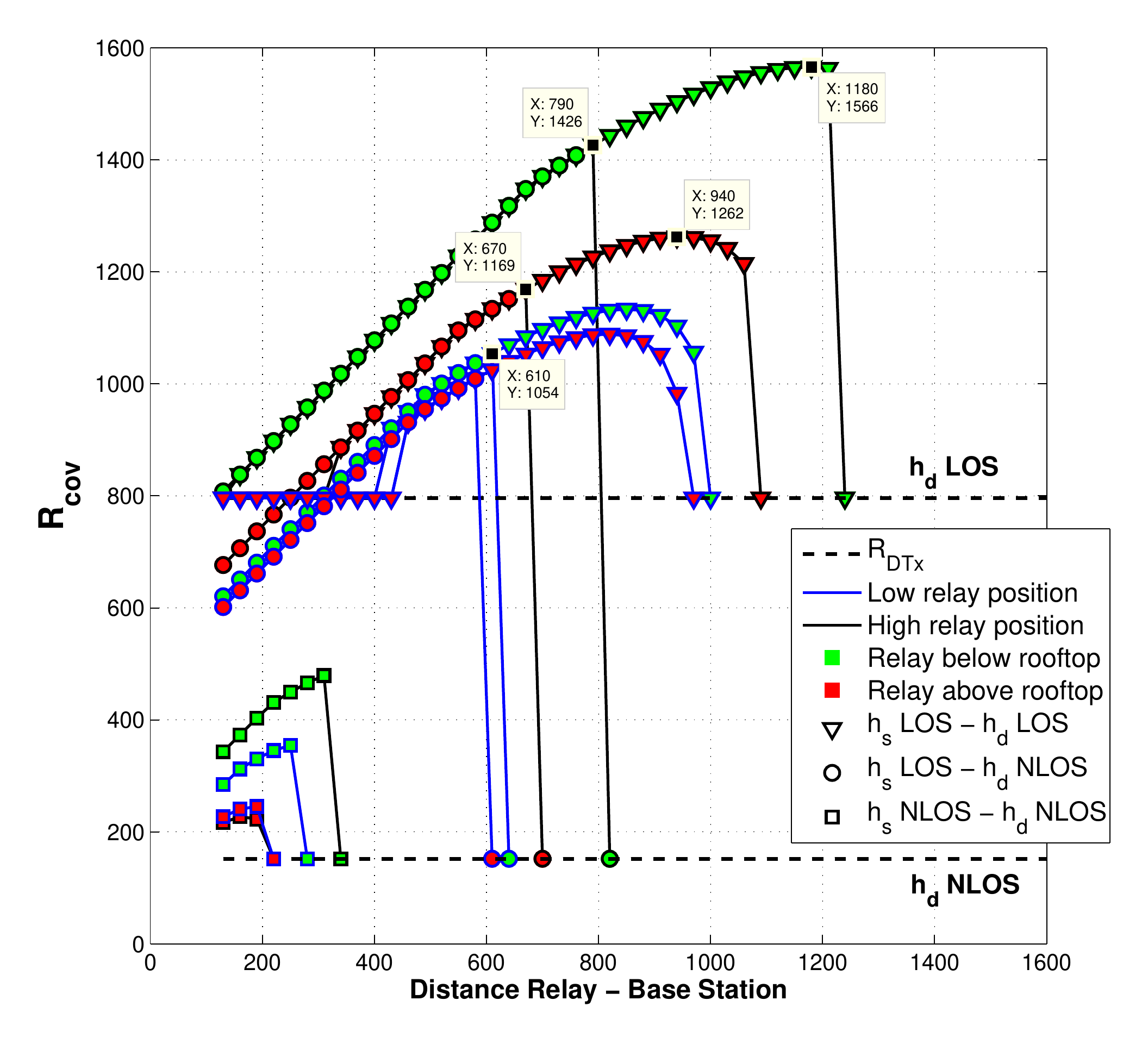}
	    \label{fig:coverage_DF}
	} 
	\subfigure[Energy gain over DTx (in dB) with $\mathsf{R}_{\text{cov}}= \mathsf{R}_{\text{DTx}}$]{
	    \includegraphics[width=0.87\columnwidth]{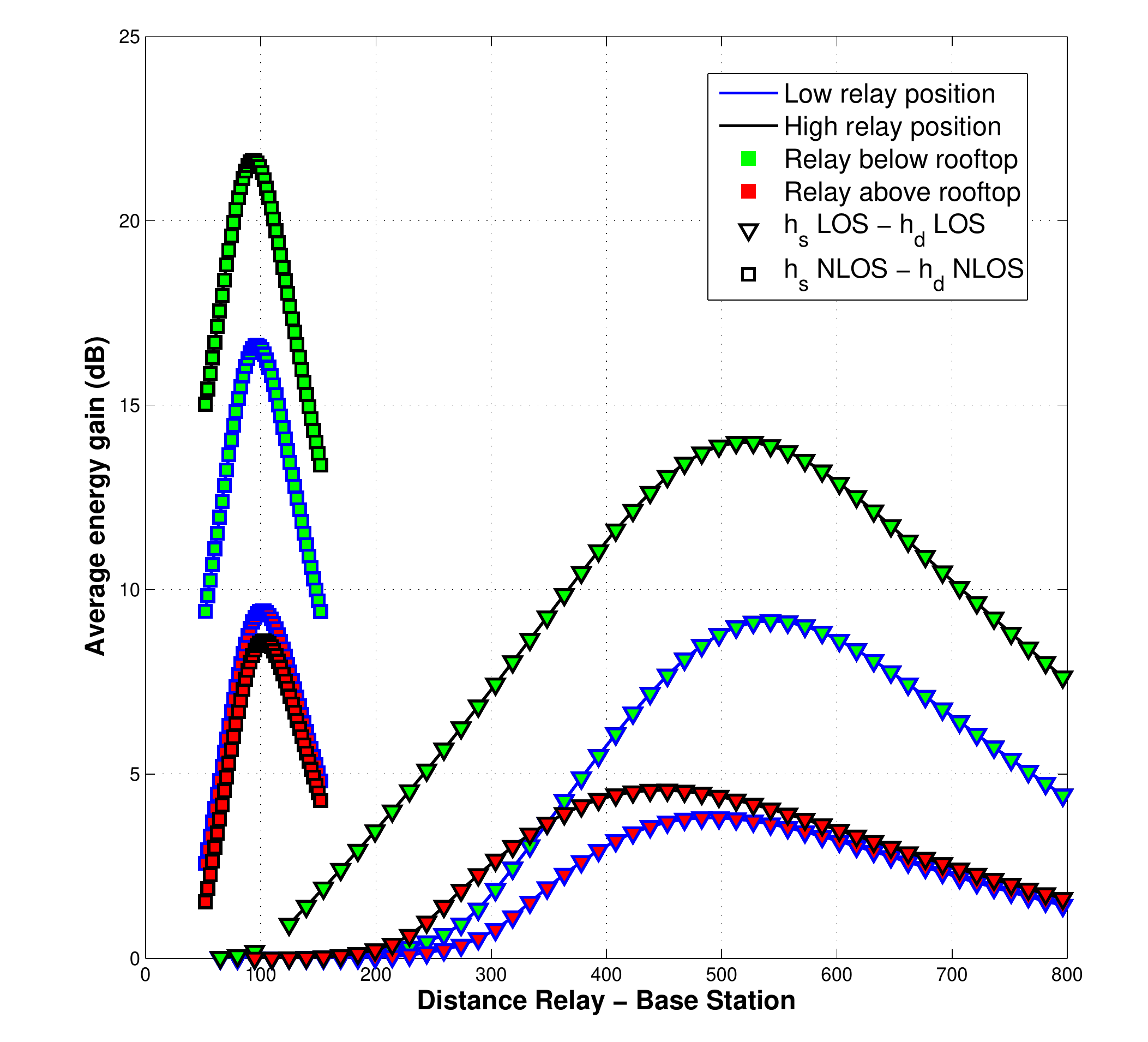}
	    \label{fig:energy_DF}
	}
	\caption{Impact of the environment using Full-DF}  
	\label{fig:Perf_FullDF} 
	}
\end{figure}

We plot in Figure \ref{fig:Perf_FullDF} the maximal cell radius and the energy gain over DTx as functions of the relay-to-base-station distance $\mathsf{D}_r$ for various relay heights and LOS conditions.
We consider the energy gain (in dB) using Full-DF compared to DTx, given that the cell coverage is not extended ($\mathsf{R}_{\text{cov}}= \mathsf{R}_{\text{DTx}}$). 
We refer to \textit{low relay location} when $\mathsf{H}_{R} = 10$m (resp. $\mathsf{H}_{R} = 20$m) and the relay is below (resp. above) rooftop. Similarly, \textit{high relay location} refers to $\mathsf{H}_{R} = 20$m (resp. $\mathsf{H}_{R} = 30$m) when the relay is below (resp. above) rooftop.
The cut-off in the curves denotes that, above the corresponding $\mathsf{D}_r$, the relay cannot cover the whole cell surface.

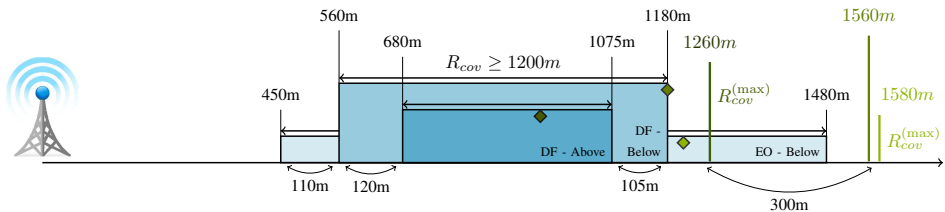
\begin{figure*}
\centering \resizebox{0.70\textwidth}{!}{%
\begin{tikzpicture}

\node [anchor=south] (BS) at (0,0) {\includegraphics[width=40pt]{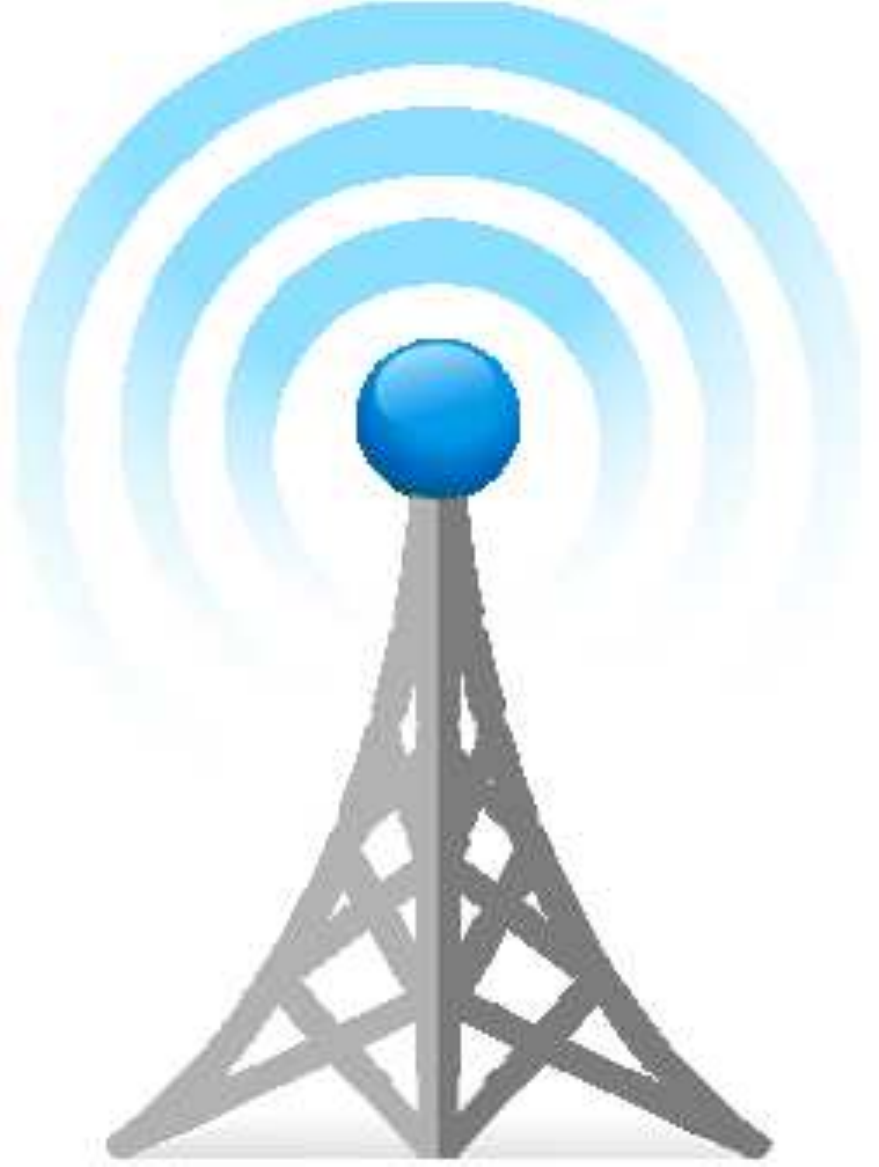}};

\node (minR_DF) at (6.8,0) {};
\node (maxR_DF) at (10.75,0) {};
\node (Rcov_DF) at (12.60,-0.1) {};

\node (minR_EO) at (5.60,0) {};
\node (maxR_EO) at (11.8,0) {};
\node (Rcov_EO) at (15.60,-0.1) {};

\node (minR_3) at (4.50,0) {};
\node (maxR_3) at (14.8,0) {};
\node (Rcov_3) at (15.80,-0.1) {};

\draw[draw=black, fill=myblue!20!white, thick, opacity=1] (minR_3)  rectangle ($(maxR_3)+(0,0.5)$);
\node[diamond, anchor=north, draw=black!80!white, fill=OliveGreen, thick,scale=0.5] at ($(12.10,0.5)$) {};

\node [anchor=south east] at (maxR_3){\color{black} \scriptsize EO - Below};

\path ($(minR_3)+(0,0.6)$) edge [<->, black, thick, above] node {$R_{cov} \geq 1200m$} ($(maxR_3)+(0,0.6)$);

\draw [-, black] ($(minR_3)+(0,0.5)$) -- ($(minR_3)+(0,1.0)$);
\node [anchor=south] at ($(minR_3)+(0,1.05)$) {\small 450m};

\draw [-, black] ($(maxR_3)+(0,0.5)$) -- ($(maxR_3)+(0,1.0)$);
\node [anchor=south] at ($(maxR_3)+(0,1.05)$) {\small 1480m};

\path (Rcov_3) edge [-, OliveGreen, very thick] node [anchor=west, black] {\color{OliveGreen} $R_{cov}^{(\max)}$} ($(Rcov_3)+(0,1.0)$);
\node [anchor=south west] at ($(Rcov_3)+(-0.1,1.15)$) {\color{OliveGreen} $1580m$};

\draw[draw=black, fill=myblue!50!white, thick, opacity=1] (minR_EO)  rectangle ($(maxR_EO)+(0,1.5)$);
\node[diamond, anchor=north, draw=black!80!white, fill=OliveGreen!70!black, thick,scale=0.5] at ($(11.80,1.5)$) {};

\node [anchor=south east, align = right] at (maxR_EO){\color{black} \scriptsize Below};
\node [anchor=south east, align = right] at ($(maxR_EO) + (0,0.4)$){\color{black} \scriptsize DF -};

\path ($(minR_EO)+(0,1.6)$) edge [<->, black, thick, above] node {$R_{cov} \geq 1200m$} ($(maxR_EO)+(0,1.6)$);

\draw [-, black] ($(minR_EO)+(0,1.5)$) -- ($(minR_EO)+(0,2.5)$);
\node [anchor=south] at ($(minR_EO)+(0,2.55)$) {\small 560m};

\draw [-, black] ($(maxR_EO)+(0,1.5)$) -- ($(maxR_EO)+(0,2.5)$);
\node [anchor=south] at ($(maxR_EO)+(0,2.55)$) {\small 1180m};

\path (Rcov_EO) edge [-, OliveGreen!70!black, very thick] node [anchor=west, black] {} ($(Rcov_EO)+(0,2.5)$);
\node [anchor=south] at ($(Rcov_EO)+(0,2.65)$) {\color{OliveGreen!70!black} $1560m$};

\draw[draw=black, fill=myblue!80!white, thick, opacity=1] (minR_DF)  rectangle ($(maxR_DF)+(0,1)$);
\node[diamond, anchor=north, draw=black!90!white, fill=OliveGreen!50!black, thick,scale=0.5] at ($(9.4,1)$) {};

\draw [myblue, very thick] (Rcov_DF) -- ($(Rcov_DF)+(0,2)$);
\node [anchor=south east] at (maxR_DF) {\color{black} \scriptsize DF - Above};

\draw [-, black] ($(minR_DF)+(0,1.0)$) -- ($(minR_DF)+(0,2.0)$);
\node [anchor=south] at ($(minR_DF)+(0,2.05)$) {\small 680m};

\draw [-, black] ($(maxR_DF)+(0,1.0)$) -- ($(maxR_DF)+(0,2.0)$);
\node [anchor=south] at ($(maxR_DF)+(0,2.05)$) {\small 1075m};

\path ($(minR_DF)+(0,1.1)$) edge [<->, black, thick, above] node {} ($(maxR_DF)+(0,1.1)$);

\path (Rcov_DF) edge [-, OliveGreen!50!black, very thick] node [anchor=south west, black] {\color{OliveGreen!50!black} $R_{cov}^{(\max)}$} ($(Rcov_DF)+(0,2.0)$);
\node [anchor=south] at ($(Rcov_DF)+(0,2.15)$) {\color{OliveGreen!30!black} $1260m$};

\path (minR_3) edge [<->, below, bend right]  node {\small 110m}  (minR_EO);
\path (minR_EO) edge [<->, below, bend right]  node {\small 120m}  (minR_DF);
\path (maxR_DF) edge [<->, below, bend right]  node {\small 105m}  (maxR_EO);
\path ($(Rcov_DF)+(-0.05,0)$) edge [<->, below, bend right]  node {\small 300m}  ($(Rcov_EO)+(0.05,0)$);

\draw [->, black, thick] (0,0) -- (17,0);

\end{tikzpicture}
}
\vspace*{-5pt}\caption{Impact of the relay height on the cell coverage}
\label{fig:relay_height_illustration}
\end{figure*}

\vspace*{-5pt}
\begin{result}
If located below the rooftop, the relay station offers not only better performance when optimally located, but it also allows a wider range of relay locations for which performance is good, even if not optimal.
\end{result} \vspace*{-5pt} \noindent
Since the relay-to-base-station link $h_r$ is not the limiting one, it is beneficial to locate the relay such that the user-to-relay link $h_s$ is stronger. Regarding energy, while a base-station-like relay (i.e. above rooftop) hardly offers 5dB gain compared to DTx, a vicinity relay (i.e. below rooftop) achieves this gain for almost all relay locations and offers up to 14dB gain at optimal location. This significant gain mostly comes from cell-edge. While all cell-edge users are transmitting with almost full power using DTx, relaying allows them to achieve similar performance to a closer user.
Regarding coverage, we can illustrate Result 1 using Figure \ref{fig:relay_height_illustration}. 
In this, the rectangles refer to the range of $\mathsf{D}_r$ for which the coverage is over 1200m. If the relay is above rooftop, like a base station, we have $R_{\text{cov}} \geq 1200m$ for $\mathsf{D}_r \in \left[680m,1075m\right]$. However, for a vicinity relay, this range is extended by more than 200m, which offers much more flexibility for a designer to locate a relay station adequately.
In Figure \ref{fig:relay_height_illustration}, the maximal coverage extension and the corresponding optimal $\mathsf{D}_r$ are respectively plotted with a green line and a green diamond. Note that the vicinity relay outperforms the base-station-like relay when optimally positioned and allows to increase the maximal coverage by 300m.

\vspace*{-5pt}\begin{result}
The system performance is strongly impacted by the effective relay height using a vicinity relay compared to a base-station-like relay. 
\end{result} \vspace*{-5pt} \noindent
Even if a vicinity relay still outperforms a base-station-like relay when being in a low position, decreasing the relay height from 20m (high relay position) to 10m (low position) causes a loss of more than 400m in the coverage and a 5dB energy loss, considering LOS condition (green triangles), as shown in Figure \ref{fig:Perf_FullDF}. Thus, with respect to both coverage and energy efficiency, the relay should be placed at the highest location possible as long as it remains below rooftop, which allows higher quality and also more probable LOS condition for the user-to-relay link than a base-station-like relay.

\vspace*{-5pt}\begin{result}
The energy-optimal distance $\mathsf{D}_r$ is different from the coverage-optimal distance.
\end{result} \vspace*{-5pt} \noindent
Regarding coverage, it is beneficial to locate the relay far from the base station. On the contrary, regarding energy consumption, the relay should be placed around the middle of the cell, so as to decrease the value for $\mathsf{D}_{\min}$ and consequently, to increase the probability of energy-efficient relaying. For example, the coverage-optimal $\mathsf{D}_r$ is 1180m for a vicinity relay with LOS conditions (green triangles) while the energy-optimal $\mathsf{D}_r$ is only 520m.

\begin{figure*}
\centering \resizebox{0.70\textwidth}{!}{%
\begin{tikzpicture}

\node [anchor=south] (BS) at (0,0) {\includegraphics[width=40pt]{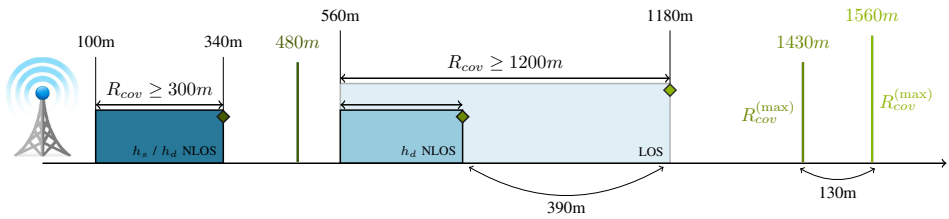}};

\node (minR_DF) at (5.6,0) {};
\node (maxR_DF) at (7.9,0) {};
\node (Rcov_DF) at (14.30,-0.1) {};

\node (minR_EO) at (5.60,0) {};
\node (maxR_EO) at (11.8,0) {};
\node (Rcov_EO) at (15.60,-0.1) {};

\node (minR_NLOS) at (1,0) {};
\node (maxR_NLOS) at (3.4,0) {};
\node (Rcov_NLOS) at (4.8,-0.1) {};

\draw[draw=black, fill=myblue!50!, thick, opacity=0.3] (minR_EO)  rectangle ($(maxR_EO)+(0,1.5)$);
\node[diamond, anchor=north, draw=black!80!white, fill=OliveGreen, thick,scale=0.5] at ($(11.80,1.5)$) {};

\path ($(minR_EO)+(0,1.6)$) edge [<->, black, thick, above] node {$R_{cov} \geq 1200m$} ($(maxR_EO)+(0,1.6)$);
\draw [myblue, very thick] (Rcov_EO) -- ($(Rcov_EO)+(0,2.5)$);

\node [anchor=south east] at (maxR_EO){\color{black} \scriptsize LOS};

\draw [-, black] ($(minR_EO)+(0,1.5)$) -- ($(minR_EO)+(0,2.5)$);
\node [anchor=south] at ($(minR_EO)+(0,2.55)$) {\small 560m};

\draw [-, black] ($(maxR_EO)+(0,1.5)$) -- ($(maxR_EO)+(0,2.5)$);
\node [anchor=south] at ($(maxR_EO)+(0,2.55)$) {\small 1180m};

\path (Rcov_EO) edge [-, OliveGreen, very thick] node [anchor=west, black] {\color{OliveGreen} $R_{cov}^{(\max)}$} ($(Rcov_EO)+(0,2.5)$);
\node [anchor=south] at ($(Rcov_EO)+(0,2.65)$) {\color{OliveGreen} $1560m$};

\draw[draw=black, fill=myblue!50!, thick, opacity=1] (minR_DF)  rectangle ($(maxR_DF)+(0,1)$);
\node[diamond, anchor=north, draw=black!90!white, fill=OliveGreen!75!black, thick,scale=0.5] at ($(7.9,1)$) {};

\draw [myblue!50!black, very thick] (Rcov_DF) -- ($(Rcov_DF)+(0,2)$);
\node [anchor=south east] at (maxR_DF) {\color{black} \scriptsize $h_d$ NLOS};

\path ($(minR_DF)+(0,1.1)$) edge [<->, black, thick, above] node {} ($(maxR_DF)+(0,1.1)$);

\path (maxR_DF) edge [<->, below, bend right]  node {\small 390m}  (maxR_EO);
\path ($(Rcov_DF)+(-0.05,0)$) edge [<->, below, bend right]  node {\small 130m}  ($(Rcov_EO)+(0.05,0)$);

\path (Rcov_DF) edge [-, OliveGreen!75!black, very thick] node [anchor=east, black] {\color{OliveGreen!75!black} $R_{cov}^{(\max)}$} ($(Rcov_DF)+(0,2.0)$);
\node [anchor=south] at ($(Rcov_DF)+(0,2.15)$) {\color{OliveGreen!75!black} $1430m$};

\draw[draw=black, fill=myblue!80!black, thick, opacity=1] (minR_NLOS)  rectangle ($(maxR_NLOS)+(0,1)$);
\node[diamond, anchor=north, draw=black!90!white, fill=OliveGreen!50!black, thick,scale=0.5] at ($(3.4,1)$) {};

\path ($(minR_NLOS)+(0,1.1)$) edge [<->, black, thick, above] node {$R_{cov} \geq 300m$} ($(maxR_NLOS)+(0,1.1)$);
\draw [myblue!80!black, very thick] (Rcov_NLOS) -- ($(Rcov_NLOS)+(0,2)$);

\node [anchor=south east] at (maxR_NLOS){\color{black} \scriptsize $h_s$ / $h_d$ NLOS};

\draw [-, black] ($(minR_NLOS)+(0,1.0)$) -- ($(minR_NLOS)+(0,2.0)$);
\node [anchor=south] at ($(minR_NLOS)+(0,2.05)$) {\small 100m};

\draw [-, black] ($(maxR_NLOS)+(0,1.0)$) -- ($(maxR_NLOS)+(0,2.0)$);
\node [anchor=south] at ($(maxR_NLOS)+(0,2.05)$) {\small 340m};

\path (Rcov_NLOS) edge [-, OliveGreen!50!black, very thick] node [anchor=east, black] {} ($(Rcov_NLOS)+(0,2.0)$);
\node [anchor=south] at ($(Rcov_NLOS)+(0,2.15)$) {\color{OliveGreen!50!black} $480m$};

\draw [->, black, thick] (0,0) -- (17,0);

\end{tikzpicture}
}
\vspace*{-5pt}\caption{Impact of the LOS conditions  on the cell coverage using Full-DF} 
\label{fig:LOS_condition_illustration}
\end{figure*}

\vspace*{-5pt}\begin{result}
The loss of LOS of $h_d$ mostly affects the range for beneficial distances $\mathsf{D}_r$, and limitedly impacts the maximum cell coverage, especially when relay is positioned in a high location but still below rooftop.
\end{result} \vspace*{-5pt} \noindent
This is illustrated in Figure \ref{fig:LOS_condition_illustration}, considering a vicinity relay.
Indeed, with LOS for the direct link, a user located in the middle of the cell can use DTx, allowing the relay station to be far from the base station. With $h_d$ NLOS, middle-cell users then require relaying to perform the transmission at same rate, meaning that the relay cannot be away from the cell center.

When both $h_s$ and $h_d$ are NLOS, performance are significantly degraded, but same comments as above can be made, i.e. vicinity relays outperform base-station-like relays but are more sensitive to the effective relay height.

\subsection{Performance evaluation with optimized coding scheme}

\begin{figure}
	{\centering 
	\subfigure[Maximal coverage extension]{
	    \includegraphics[width=0.87\columnwidth]{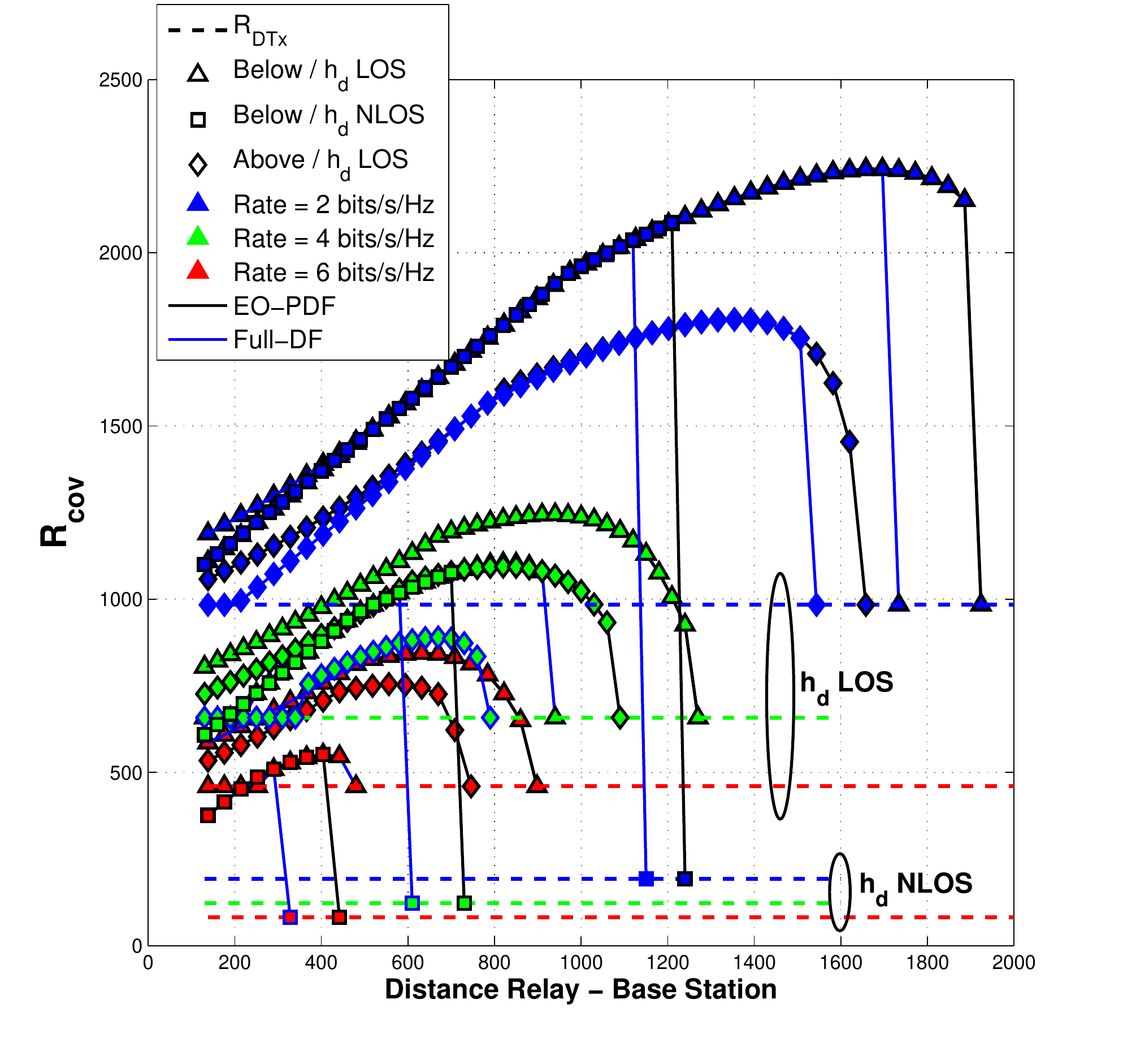}
	    \label{fig:with_EO_PDF_c}
	} 
	\subfigure[Energy consumption with $\mathsf{R}_{\text{cov}}= \mathsf{R}_{\text{cov}}^{(\text{DF})}$]{
	    \includegraphics[width=0.87\columnwidth]{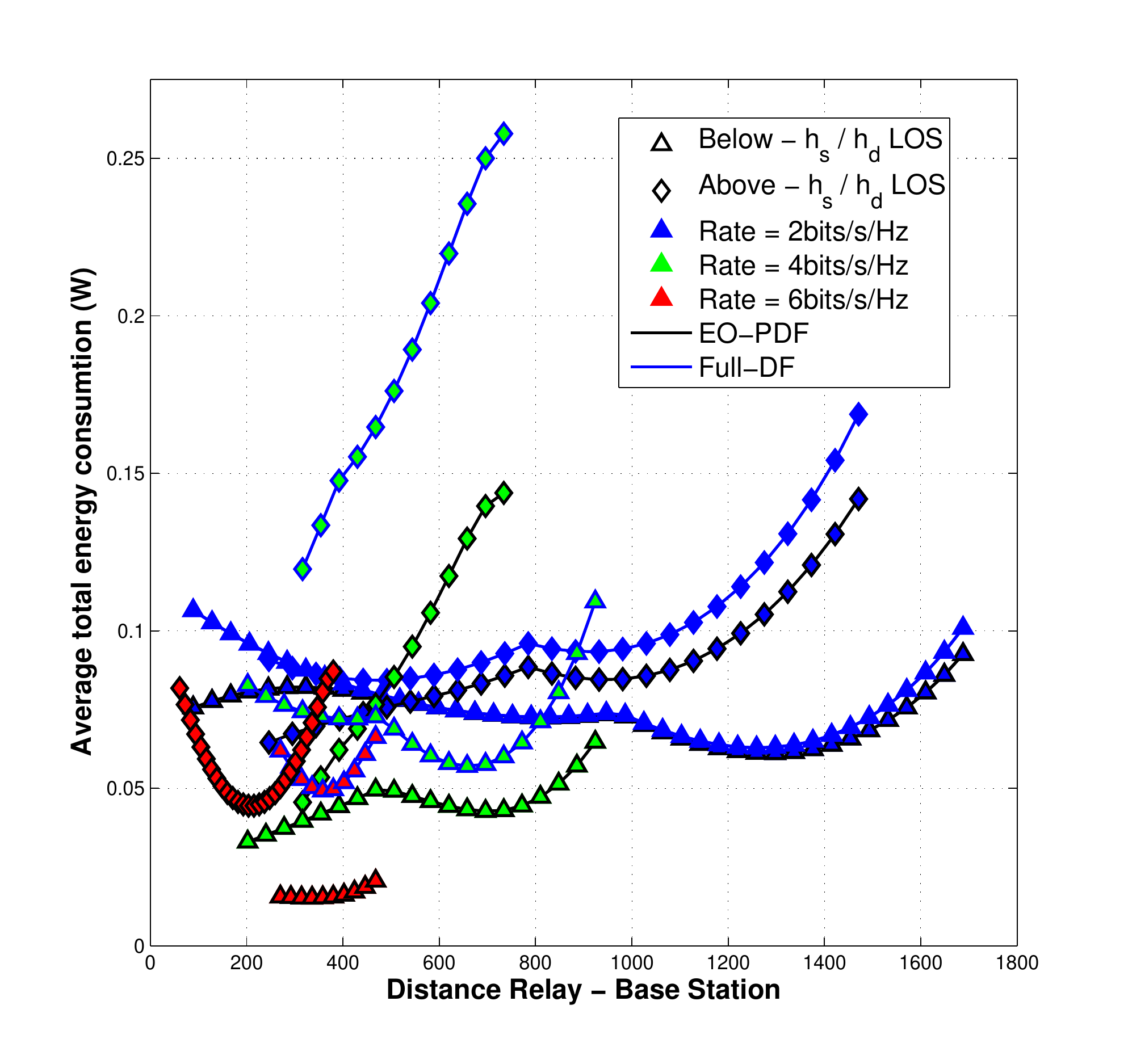}
	    \label{fig:with_EO_PDF_e}
	}
	\caption{Comparison between Full-DF and EO-PDF} 
	\label{fig:with_EO_PDF} 
	}
\end{figure}

So far, we have analyzed the system performance for the practical Full-DF scheme. We now compare the results with the upper-bound performance obtained by using the energy-optimal scheme EO-PDF.
We plot in Figure \ref{fig:with_EO_PDF} the maximum coverage radius and the average total energy consumed by each scheme as functions of the relay-to-base-station distance. We consider various environment conditions and several user rates. The energy consumption is computed with $\mathsf{R}_{\text{cov}}= \mathsf{R}_{\text{cov}}^{(\text{DF})}$.
Also note that it consists of two parts due to the hexagonal cell constraint since the coverage radius is written as a minimum of two terms, as shown in Eq. \eqref{eq:R_cov}. The first part of the curve corresponds to $\mathsf{X}_{\max} + \mathsf{R}_{\max}$ being the minimum.

\vspace*{-5pt}\begin{result}
With regards to both coverage and energy, the EO-PDF performs significantly better than Full-DF for high rates and under weaker channel conditions, i.e. when the direct link is NLOS (plotted with squares) or when the relay is above rooftop (diamonds). 
\end{result} \vspace*{-5pt} \noindent
Given good channel conditions (plotted with triangles), i.e. vicinity relay and LOS conditions, Full-DF performs close to the EO-PDF scheme for low rates ($\mathcal{R} \leq 2$bit/s/Hz), with regards to both coverage and energy. However, performances of Full-DF are rapidly degraded when the user rate increases. As illustrated in Figure \ref{fig:relay_height_illustration} for $\mathcal{R}=3$bit/s/Hz, the maximum cell coverage obtained with Full-DF remains close to the coverage given by the EO-PDF, but the range of beneficial relay locations for which $R_{\text{cov}} \geq 1200m$ is far smaller. Also note that high rates ($\mathcal{R}=6$bit/s/Hz) are only achievable with base-station-like relay (i.e. above rooftop) by using the EO-PDF, and with vicinity relays, the coverage using Full-DF is only 65\% of the one obtained using EO-PDF.

Same comments can be made with regards to energy efficiency.
For example, when $\mathsf{D}_r=650$m with $\mathcal{R}=4$bits/s/Hz, the energy loss of Full-DF is around 1.2dB compared to EO-PDF. This loss is relatively low compared to the increase in complexity necessary to implement EO-PDF. However, the performance of Full-DF is significantly degraded when the relay is not optimally positioned. If the relay is closer to the base station ($\mathsf{D}_r \leq 400$m), between 2 and 4dB energy loss is observed. For $\mathcal{R}=6$bits/s/Hz, this loss attains 6dB.

\subsection{Analysis of downlink performance}

\begin{figure}
	\centering
	\includegraphics[width=0.87\columnwidth]{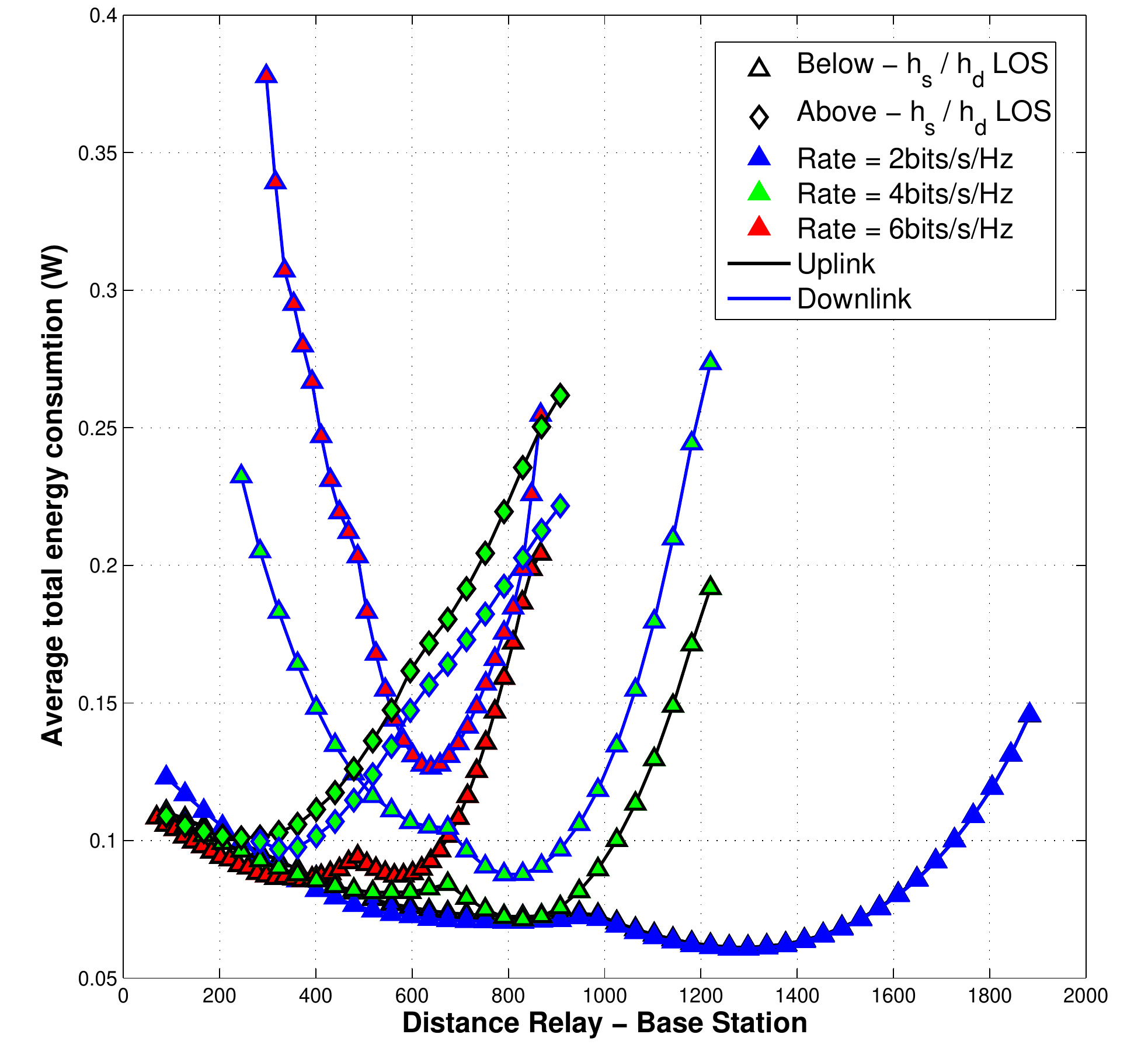} 
	\caption{Downlink and uplink performance using EO-PDF}  
	\label{fig:downlink} 
\end{figure}

So far, we have analyzed the performance of relaying for uplink transmissions, we know investigate downlink. Due to the symmetry of the scheme, Full-DF has the same performance for both. We thus focus only on the EO-PDF scheme.
In Figure \ref{fig:downlink}, we plot the average total energy consumed by the EO-PDF scheme given the cell coverage obtain for uplink ($\mathsf{R}_{\text{cov}}= \mathsf{R}_{\text{cov}}^{(\text{EO,UL})}$).
\vspace*{-5pt}\begin{result}
Vicinity relays still outperforms base-station-like relays. Interestingly, EO-PDF is more energy-efficient for uplink transmissions than for downlink.
\end{result} \vspace*{-5pt} \noindent
This is due to equal time division which is suboptimal, especially when the relay is located above rooftop. In this case, the link from base station to relay station is very strong. Thus, the EO-PDF scheme tends to relay the whole message, using limited rate splitting ($m \sim m_r$), and consumes most of energy in the second phase to take advantage of the beamforming gain between the base station and the relay. On the contrary, locating the relay station below rooftop allows more balance between $h_s$ and $h_r$. This implies better use of rate splitting which is more suitable for equal time division.

%

\subsection{Trade-off between coverage extension and energy efficiency}

Here, we refine the analysis of the energy consumption vs.
the covered area.
Indeed, increasing the coverage radius implies that far-users are included within the cell. Those far-users consuming a lot of energy due to distance, the energy efficiency is thus decreased. However, extending the cell coverage reduces the number of base stations required to cover a given area, which is cost-efficient for the network deployment.
Consequently, a network designer may opt for better energy efficiency regardless of the coverage, or may prefer to insure a larger coverage even if more energy is consumed in average. 
This constitutes a fundamental trade-off on relay deployment.

\begin{figure}
	{\centering 
	\subfigure[Region for optimal trade-off]{
	    \includegraphics[width=0.87\columnwidth]{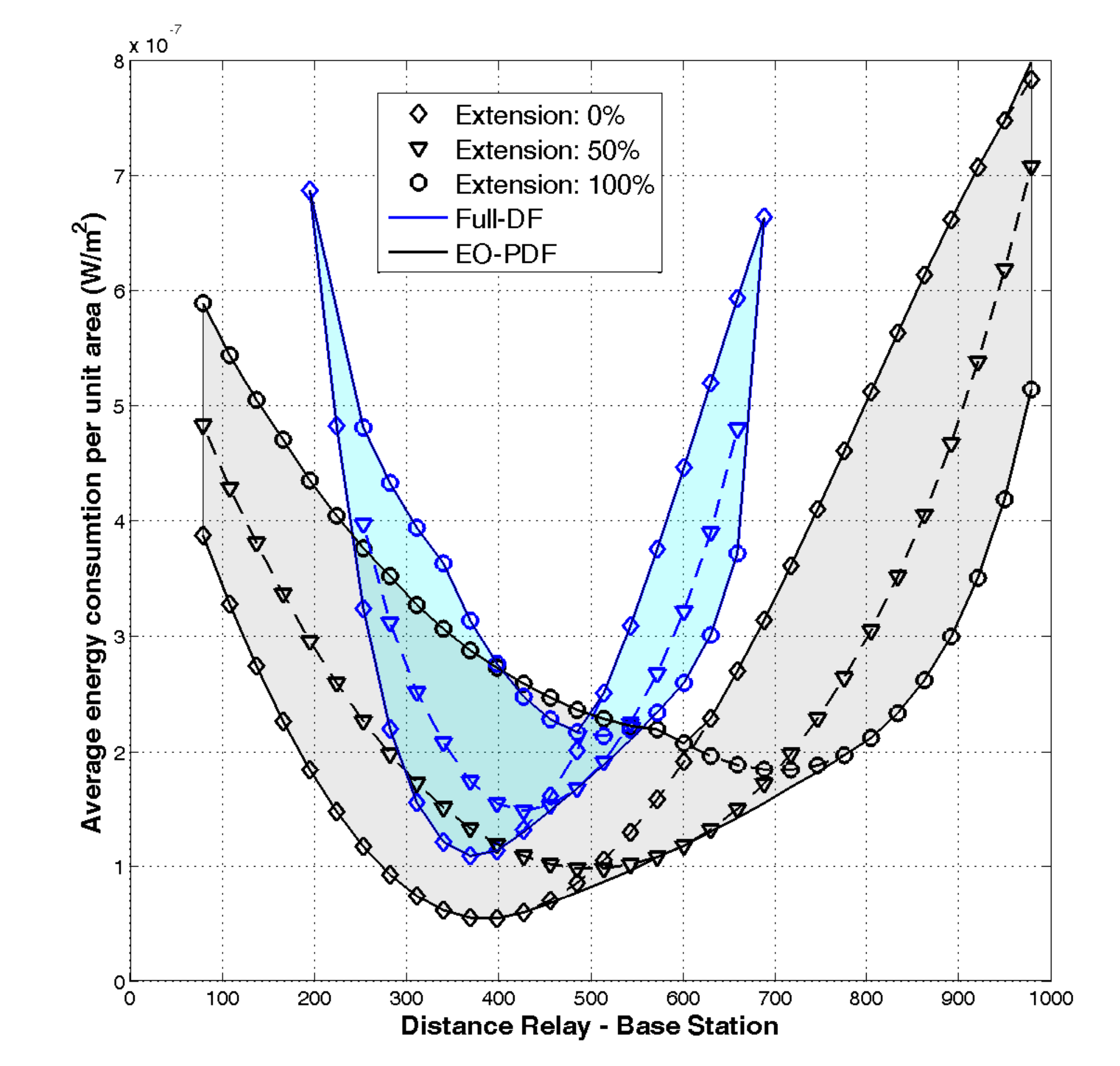}
	    \label{fig:trade_off_energy}
	} 
	\subfigure[Deployment cost]{
	    \includegraphics[width=0.87\columnwidth]{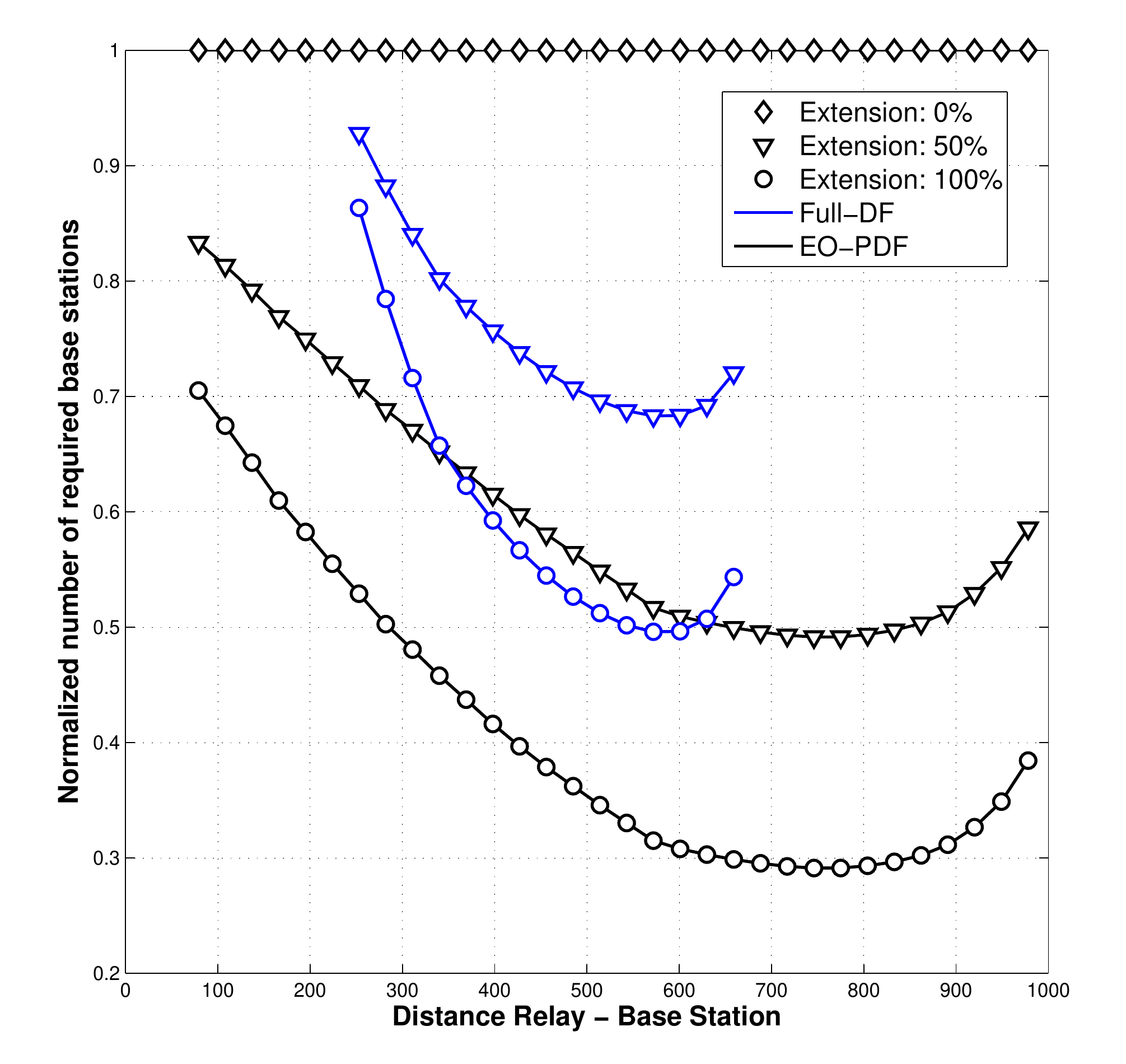}
	    \label{fig:trade_off_bs}
	}
	\caption{Trade-off between coverage extension and energy consumption, with $\mathcal{R}= 5$bits/s/Hz}  
	\label{fig:trade_off}
	}
\end{figure}

We thus propose to simulate performance based on the average energy consumption per unit area. As defined in \cite{correia2010}, this is the ratio of the average energy consumption over the coverage area $\frac{\mathbb{E}\left[E\right]}{\mathcal{A}_{\text{sector}}}$, with $\mathbb{E}\left[E\right]$ and $\mathcal{A}_{\text{sector}}$ as expressed in Eq. \eqref{eq:energy} and \eqref{eq:aire_} respectively.
We plot in Figure \ref{fig:trade_off_energy} the average energy consumption per unit area considering several coverage radius, expressed as
$R_{\text{cov}} = R_{\text{DTx}} + \beta \left( R_{\text{cov}}^{\text{(max)}}- R_{\text{DTx}} \right)$,
where $\beta$ refers to the coverage extension in percentage and $R_{\text{cov}}^{\text{(max)}}$ to the maximum coverage radius as plotted in Figure \ref{fig:with_EO_PDF_c}.
By plotting all curves given by $\beta \in \left[ 0,100\% \right]$, we obtain the shaded area, which corresponds to the region of optimal trade-off. For any set $(\frac{\mathbb{E}\left[E\right]}{\mathcal{A}_{\text{sector}}},\mathsf{D}_r)$ taken in this region, there exists a coverage $R_{\text{cov}}$ for which this energy per unit area is optimal with this relay-to-base-station distance $\mathsf{D}_r$.
In Figure \ref{fig:trade_off_bs}, we plot the ratio of the area covered by direct transmission over the extended area $\frac{\mathcal{A}_{\text{DTx}}}{\mathcal{A}_{\text{RTx}}}$. It refers to the normalized number of required base stations, i.e. the deployment cost, and is directly linked to the maximal coverage extension, as plotted in Figure \ref{fig:with_EO_PDF_c}. For example, at $\mathsf{D}_r=$600m, using EO-PDF with maximal coverage extension requires only 30\% of the base stations needed to cover the same area without cell extension.

The lower bound of the shaded area in Figure \ref{fig:trade_off_energy} gives the minimal energy consumed per unit area as a function of $\mathsf{D}_r$ and does not refer to a single coverage extension $\beta$.
We observe once again that the EO-PDF scheme offers better performances than Full-DF both at the optimal distance $\mathsf{D}_r=375$m and around this optimum. However, when maximum coverage extension ($\beta=100\%$) is considered, Full-DF outperforms the energy-optimized scheme EO-PDF for some values of $\mathsf{D}_r$. As explained earlier, far-users that consume large amount of energy are included within the cell by using EO-PDF, thus illustrating the need for defining a trade-off between energy and coverage.

\vspace*{-5pt}\begin{result}
The minimal average energy consumption per unit area is achieved when the cell coverage is not extended, i.e. when only energy efficiency is considered.
\end{result} \vspace*{-5pt} \noindent
This minimum is attained with $\mathsf{D}_r=375$m for both Full-DF and EO-PDF. Up to this distance $\mathsf{D}_r$, no coverage extension ($\beta=0\%$) is more energy-efficient than extended cell. Above this distance, coverage extension has to be considered for better energy efficiency. Note that, when the relay is far from the base station (over 800m for EO-PDF and 600m for Full-DF), reducing the coverage extension is no more energy-efficient.

\vspace*{-5pt}\begin{result}
The minimal deployment cost, i.e. the minimal required number of base stations, is achieved when the relay is far from the base station, but shows in return severe loss regarding the energy consumed per unit area.
\end{result} \vspace*{-5pt} \noindent
Between the energy-optimal relay location ($\mathsf{D}_r=375$m) and the cost-optimal location ($\mathsf{D}_r=$750m for EO-PDF and 575m for Full-DF), the energy loss, compared to the lower-bound of the shaded area, is equal to 5.2dB for EO-PDF and 3.3dB for Full-DF.

\vspace*{-5pt}\begin{result}
There exists a relay-to-base-station distance $\mathsf{D}_r$ for which extending or reducing the cell coverage has minimum impact on the energy consumed per unit area.
\end{result} \vspace*{-5pt} \noindent
This impact is illustrated in Figure \ref{fig:trade_off_energy} by the thickness of the shaded area at a given $\mathsf{D}_r$. For example, at the energy-optimal $\mathsf{D}_r=$375m, extending the cell coverage from $R_{\text{DTx}}$ (Extension: 0\%) to $R_{\text{cov}}^{\text{(max)}}$ (Extension: 100\%) induces an significant energy loss of 7dB for EO-PDF. This loss is reduced to 2.5dB when $\mathsf{D}_r=$600m, distance for which the deployment cost is only 31\% of the cost of non-extended cells. 

To summarize results, energy-efficient relay placement should be jointly considered with cell coverage extension, and a moderate extension provides better performance in terms of energy. However, the optimal trade-off is, to our mind, 
neither the energy-optimal nor the cost-optimal relay location, but rather the location for which modifying the cell coverage has minimum impact on the energy consumed per unit area. This is particularly suitable for heterogeneous networks. 
Indeed, the deployment of unplanned nodes, especially at cell edge, implies that users are served by those
nodes rather than by the base station, which virtually reduces the coverage radius of the macrocell.
Relay deployment cannot be considered to be energy-efficient only for a given coverage and showing up a 7dB loss as soon as a pico- or femtocell is added, or reversely, switched off.
This optimal trade-off is also relevant in networks where dynamic resource allocations and sleep mode operations enable the change of cell coverage as a function of the cell load and desired user rates.

\section{Conclusion}
\label{sec:conclusion}
We have investigated relay placement for noise-limited urban cells and considered both cell coverage and energy efficiency.
Using an approach complementary to designing algorithms for optimal placement existing in the literature, we analyzed how the propagation environment, user rate and relay coding scheme affect the choice of the relay location and the network performance. 
To do so, we proposed a geometrical model for evaluating energy efficiency and coverage extension, and highlighted that a trade-off exists between them.

Two options can be deduced from this work to efficiently deploy relays in a cell. On the one hand, a network designer may choose to keep terminals simple and use two-hop routing. In this case, attention has to be paid on carefully positioning the relay station, so as to provide close to optimal coverage or energy gain. On the other hand, this designer may be constrained by the cell topology and restricted in potential relay locations. Then, more complex coding schemes, such as partial decode-forward, should be considered to fully benefit from the relay station.
With regards to heterogeneous networks, 
the relay should be located such that increasing the cell coverage has a minimal impact on energy efficiency.

\appendices

\section{Considered path-loss models}
\label{App:pathloss}

In this paper, we propose to model the path-loss depending on the relay height and base our model on the the WINNER II project \cite{winner}. 
We consider an urban environment where the user is located outdoors at street level and the base station is clearly above surrounding buildings.
Both the base station and the relay station are fixed. As recommended by 802.16j/m and 3GPP specifications \cite{hr_LOS}, the relay station establishes a high quality link with the base station and LOS is assumed for this channel. Since the relay height is allowed to vary from below to above rooftop, the propagation environment for $h_r$ and $h_s$ changes notably. However, the WINNER II project does not provide a continuous model as a function of the relay height. We will therefore consider the two situations summarized in Table \ref{sim_pathloss}.
Furthermore, we assume that the terminal are sufficiently robust against small-scale environment parameters, such as multipath components.

\begin{table}
\centering  \begin{tabular}{|c|c|c|} 
\hline
& Vicinity Relay ($\mathsf{H}_R \leq 20$m) & Base-station-like Relay ($\mathsf{H}_R \geq 20$m)\\
\hline
$h_s$ & B1 (LOS / NLOS) &  C2 (LOS / NLOS)\\
$h_r$ & B5c LOS & Free space (FS) \\
$h_d$ &\multicolumn{2}{|c|}{C2 (LOS / NLOS)} \\
\hline
\end{tabular}
\caption{Considered WINNER II scenarios} 
\label{sim_pathloss}
\end{table}

\section{Proof of Probability of Relaying in Eq. \eqref{Eq:Prob_RTx}}
\label{appendix:P_RTx}

\begin{figure}
\centering \includegraphics[width=0.87\columnwidth]{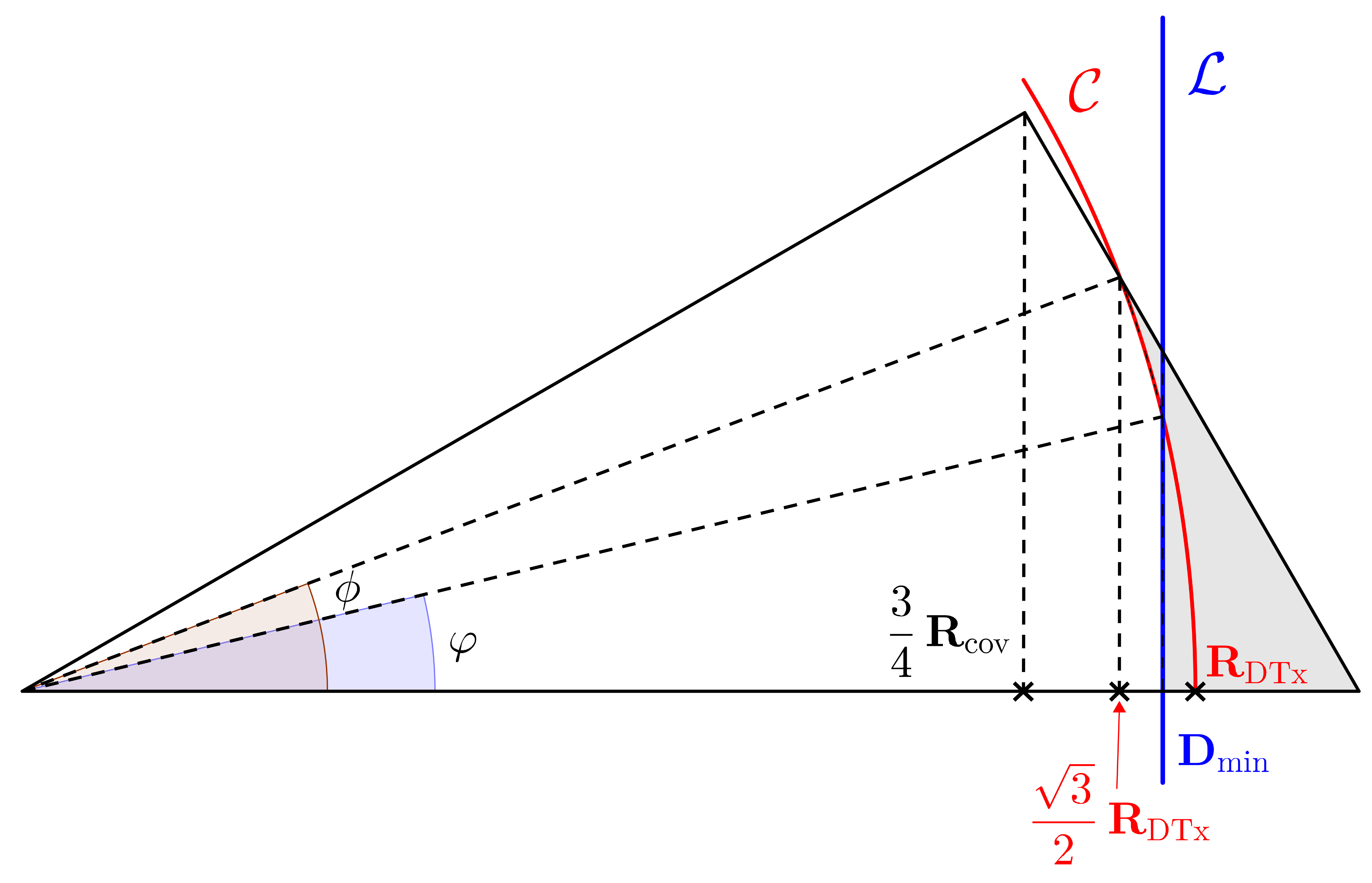}
\caption{Illustration of the Relay Efficiency Area for a half cell sector} 
\label{fig:Proof_PRTx} 
\end{figure}

The expression of Eq. \eqref{Eq:Prob_RTx} is obtained using geometrical principles. $\mathbb{P}_{\text{RTx}}$  is equal to the surface of the REA divided by the surface of the cell. We can reduce the analysis to one cell sector and rather focus on the surface of $\overline{\text{REA}}$, where DTx is more energy-efficient than RTx.
We denote $\mathcal{L}$ the strait line of equation $ x=\mathsf{D}_{\min}$ and $\mathcal{C}$ the circle centred at the base station and of radius $\mathsf{R}_{\text{DTx}}$.
This is illustrated in Figure \ref{fig:Proof_PRTx}.

First, we focus on the angles $\varphi$ and $\phi$.
When $\frac{\sqrt{3}}{2}\mathsf{R}_{\text{DTx}} \leq  \frac{3}{4}\mathsf{R}_{\text{cov}}$, the cell-edge is never limiting and $\overline{\text{REA}}$ is upper-bound by either $\mathcal{L}$ or $\mathcal{C}$ (left part of the hexagonal cell sector). This gives $\phi = \frac{\pi}{6}$.
Furthermore, if $\mathsf{D}_{\min} \leq \cos \left( \frac{\pi}{6}\right)\mathsf{R}_{\text{DTx}}$, the set of all user positions within the sector for which $x < \mathsf{D}_{\min}$ is strictly included in the sphere $\mathcal{C}\left(0, \mathsf{R}_{\text{DTx}}\right)$. Therefore, $\overline{\text{REA}}$ is upper-bounded by $\mathcal{L}$ only and $\varphi = \phi$. 
Otherwise, both $\mathcal{L}$ and $\mathcal{C}$ upper-bound $\overline{\text{REA}}$ and intersect at $\left(\mathsf{R}_{\text{DTx}},\varphi \right)$, with $\varphi = \arccos \left(\frac{\mathsf{D}_{\min}}{\mathsf{R}_{\text{DTx}}}\right)$.
Second, when $\frac{3}{4}\mathsf{R}_{\text{cov}} \leq \frac{\sqrt{3}}{2}\mathsf{R}_{\text{DTx}}$ (right part of the hexagonal cell sector), the cell edge can be limiting. We denote $\mathsf{X}$ the x-coordinate of the intersection of $\mathcal{C}$ with the cell edge.
If $\mathsf{D}_{\min} \leq \frac{3}{4}\mathsf{R}_{\text{cov}}$, as before, $\overline{\text{REA}}$ is upper-bounded by $\mathcal{L}$ only and $\varphi = \phi$.
If $\frac{3}{4}\mathsf{R}_{\text{cov}} \leq \mathsf{D}_{\min} \leq \mathsf{X}$, $\overline{\text{REA}}$ is upper-bounded by both $\mathcal{L}$ and the cell edge, which gives $\varphi = \phi$.
Finally, if $\mathsf{X} \leq \mathsf{D}_{\min} $, $\overline{\text{REA}}$ is upper-bounded by $\mathcal{L}$, $\mathcal{C}$ and the cell edge, and angles are computed using geometrical properties.
This gives the expressions of $\varphi$ and $\phi$ as given in Eq. \eqref{eq:angles}.

We can now deduce the surface of $\overline{\text{REA}}$. To do so, we decompose this surface into elementary geometrical shapes, whose surface can be easily computed.
For $\theta \in \left[0, \varphi\right]$, $\overline{\text{REA}}$ is upper-bounded by $\mathcal{L}$ and its surface reduces to a triangle for this range of angles.
For $\theta \in \left[\varphi, \phi \right]$, $\overline{\text{REA}}$ is upper-bounded by $\mathcal{C}$. The related surface here is a portion of sphere. Finally, 
when $\phi < \frac{\pi}{6}$, $\overline{\text{REA}}$ is upper-bounded by the cell edge for $\theta \in \left[\phi, \frac{\pi}{6} \right]$. This also corresponds to the surface of a triangle.
Using, this decomposition, we get Eq. \eqref{Eq:Prob_RTx}.

\bibliographystyle{IEEEtranN}
{\footnotesize 
\bibliography{RefRelayPosition_v1}
}
\fontsize{10}{10}
\selectfont
\begin{IEEEbiography}[{\includegraphics[width=1in,height=1.25in,clip,keepaspectratio]{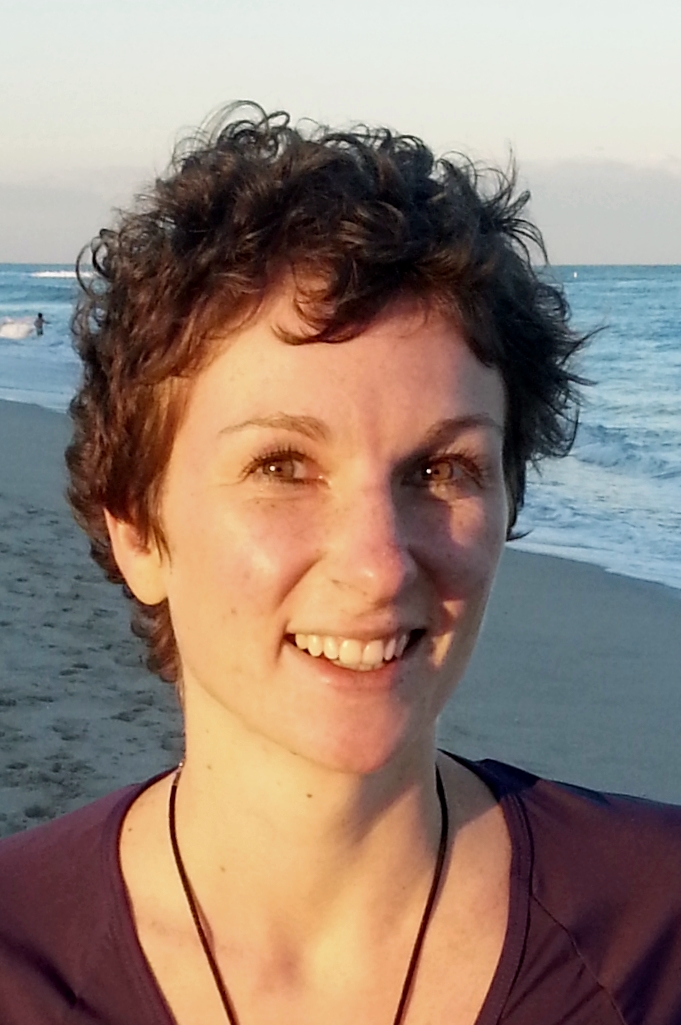}}]{Fanny Parzysz}
received the M.Sc. degree in digital communications and cellular networks from Telecom ParisTech (ENST), Paris, France, in 2009. She is currently pursuing the Ph.D. degree at the LACIME Laboratory, \'{E}cole de Technologie Sup\'{e}rieure, Montreal, Canada. Her Ph.D. is part of the NSERC-Ultra Electronics Industrial Chair in Wireless Emergency and Tactical Communication. 
Her research interest covers technologies and models for next generation relay-aided wireless networks, with a particular focus on information theory, energy-efficient coding and resource allocation.
\end{IEEEbiography}
\vfill
\newpage
\begin{IEEEbiography}[{\includegraphics[width=1in,height=1.25in,clip,keepaspectratio]{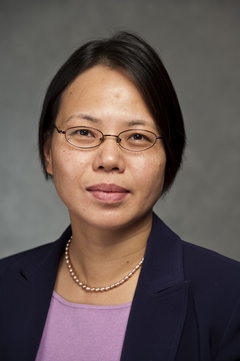}}]{Mai Vu}
received a PhD degree in Electrical Engineering from Stanford University after having an MSE degree in Electrical Engineering from the University of Melbourne and a bachelor degree in Computer Systems Engineering from RMIT, Australia. Between 2006-2008, she worked as a lecturer and researcher at the School of Engineering and Applied Sciences, Harvard University. During 2009-2012, she was an assistant professor in Electrical and Computer Engineering at McGill University. Since January 2013, she has been an associate professor in the department of Electrical and Computer Engineering at Tufts University.

Dr. Vu conducts research in the general areas of wireless systems, signal processing, and network communications. She has published extensively on cooperative and cognitive communications, relay networks, MIMO capacity and precoding. Dr. Vu has served on the technical program committee of numerous IEEE conferences and is
currently an editor for the IEEE Transactions on Wireless Communications. She is a senior member of the IEEE.
\end{IEEEbiography}

\begin{IEEEbiography}[{\includegraphics[width=1in,height=1.25in,clip,keepaspectratio]{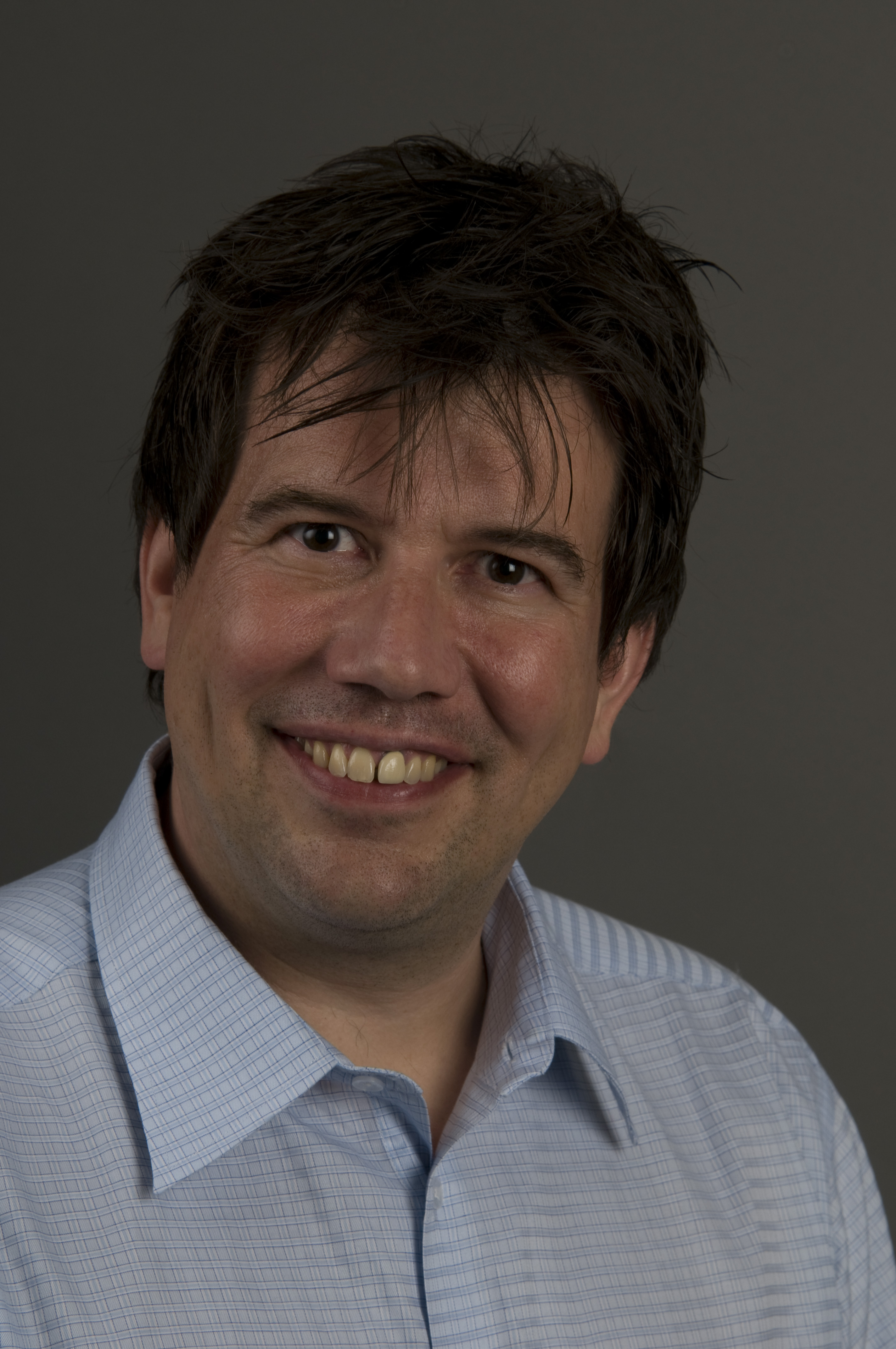}}]{Fran\c{c}ois Gagnon}
received the B.Eng. and Ph.D. degrees in electrical engineering from  \'{E}cole Polytechnique de Montr\'{e}al, Montreal, Quebec, Canada. Since 1991, he has been a Professor with the Department of Electrical Engineering,  \'{E}cole de Technologie Sup\'{e}rieure, Montreal, Quebec, Canada. He chaired the department from 1999 to 2001, and is now the holder of the NSERC Ultra Electronics Chair, Wireless Emergency and Tactical Communication, at the same university. His research interest covers wireless high-speed communications, modulation, coding, high-speed DSP implementations, and military point-to-point communications. He has been very involved in the creation of the new generation of high-capacity line of-sight military radios offered by the Canadian Marconi Corporation, which is now Ultra Electronics Tactical Communication Systems. The company has received, for this product, a “Coin of Excellence” from the U.S. Army for performance and reliability. Prof. Gagnon was awarded the 2008 NSERC Synergy Award for the fruitful and long lasting collaboration with Ultra Electronics TCS.
\end{IEEEbiography}
\vfill

\end{document}